\documentclass[12pt,draftclsnofoot,onecolumn]{IEEEtran}\newcommand{\TWOCOLUMN}[1]{}
\usepackage[pdftex]{graphicx}
\usepackage{epstopdf}

\usepackage{cite,setspace,url}
\usepackage{latexsym}
\ifCLASSINFOpdf
\else
\fi
\usepackage{verbatim}

\usepackage[cmex10]{amsmath}
\usepackage{amssymb}
\usepackage{fixltx2e}
\usepackage{texdef2015}
\usepackage[normalem]{ulem}
\usepackage[makeroom]{cancel}
\allowdisplaybreaks 

\newcommand{\Tcal}{\mathcal{T}}
\newcommand{\Ical}{\mathcal{I}}
\newcommand{\age}{\Delta}

\newcommand{\interDep}{D}
\newcommand{\interDblock}{B}
\newcommand{\Spre}{T}
\newcommand{\jlong}[1][j]{L_{#1}}
\newcommand{\jbrief}[1][j]{B_{#1}}
\newcommand{\Tavg}[2][\mathcal{T}]{\left\langle{#2}\right\rangle_{#1}}

\newcommand{\Lcal}{\mathcal{L}}
\newcommand{\Qcal}{\mathcal{Q}}
\newcommand{\Acal}{\mathcal{A}}
\newcommand{\Bcal}{\mathcal{B}}
\newcommand{\R}{\mathbb{R}}
\newcommand{\dq}[2][q]{\delta_{#2,#1}}
\newcommand{\dqt}[2][q(t)]{\delta_{#2,#1}}
\newcommand{\ql}{q_l}

\newcommand{\laml}[1][l]{\lambda^{(#1)}}

\newcommand{\psim}[2][j]{\psi_{#2#1}}
\newcommand{\Lpsim}[2][j]{L\psi_{#2#1}}
\newcommand{\Lamm}[2][j]{\Lambda_{#2#1}}
\newcommand{\cvec}[1]{\Bracket{#1}}

\newcommand{\smvec}[1]{\left[\begin{smallmatrix} #1\end{smallmatrix}\right]}
\newcommand{\rowvec}[1]{[\begin{matrix} #1\end{matrix}]}
\newcommand{\rvec}[1]{\rowvec{#1}}

\newcommand{\piv}{\text{\boldmath{$\pi$}}}
\newcommand{\pivbar}{\text{\boldmath{$\bar{\pi}$}}}
\newcommand{\pibar}{\bar{\pi}}
\newcommand{\vvbar}{\bar{\vv}}

\newcommand{\agesum}{\age_{\Sigma}}
\newcommand{\zerov}[1][\mbox{}]{\mathbf{0}_{#1}}
\newcommand{\onev}[1][n]{{\mathbf{1}}_{#1}}

\renewcommand{\vec}[1]{\begin{bmatrix} #1\end{bmatrix}}
\renewcommand{\bv}{\mathbf{b}}
\newcommand{\range}[2][0]{#1\,{:}\,#2}

\usepackage{tikz}
\usetikzlibrary{automata,arrows,positioning,calc,shapes.multipart,chains}

\newcommand{\OthersRate}[1]{\ensuremath \lambda_{-#1}}
\newcommand{\OthersLoad}[1]{\ensuremath \rho_{-#1}}
\newcommand{\Nage}[1][N]{\age^{#1}}
\newcommand{\NageF}[1][N]{\Nage[#1]_F}
\newcommand{\NageW}[1][N]{\Nage[#1]_W}
\newcommand{\NageS}[1][N]{\Nage[#1]_S}

\newcommand{\aw}{\alpha_W}

\usepackage{ifthen}

\newcommand{\Others}[1]{\ifthenelse{\equal{#1}{1}}{2}{1}}

\newcounter{lettercount}
\newenvironment{letterate}{\begin{list}%
{(\alph{lettercount})}{\usecounter{lettercount}}}{\end{list}}
\begin{document}

\setboolean{mytwocolumn}{false}

\author{Roy~D.~Yates
        and~Sanjit~K.~Kaul%
\thanks{Roy Yates is with WINLAB and the ECE Department, Rutgers University, NJ, USA, e-mail: ryates@winlab.rutgers.edu.}
\thanks{Sanjit Kaul is with Wireless Systems Lab, IIIT-Delhi, India, e-mail: skkaul@iiitd.ac.in.}
\thanks{This work was presented in part at the 2012 Conference on Information Sciences and Systems and the 2012 IEEE International Symposium on Information Theory. This work was supported by NSF award CCF-1422988.}}


%

\title{The Age of Information: Real-Time Status Updating by Multiple Sources}

\maketitle
\begin{abstract}
We examine multiple independent sources providing status updates to a monitor through simple queues. We formulate an Age of Information (AoI) timeliness metric and derive a general result for the AoI that is applicable to a wide variety of multiple source service systems. For first-come first-served and two types of last-come first-served systems with Poisson arrivals and exponential service times, we find the region of feasible average status ages for multiple updating sources. We then use these results to characterize how a service facility can be shared among multiple updating sources. A new simplified technique for evaluating the AoI in finite-state continuous-time queueing systems is also derived. Based on stochastic hybrid systems, this method makes AoI evaluation  to be comparable in complexity to finding the stationary distribution of a finite-state Markov chain. 

\end{abstract}

%
\section{Introduction}

Increasingly ubiquitous connectivity to communication networks and availability of portable devices have engendered a host of applications in which sources -- people and environmental sensors -- send updates of their status to interested recipients.
These include news and weather reports and updates by individuals on Twitter about what is keeping them busy, updates by environmental sensors~\cite{mainwaring_wireless_2002}, and vehicular  status (position, velocity, acceleration) updates that can assist drivers of nearby vehicles in an intelligent transportation system~\cite{papadimitratos_vehicular_2009}.
These applications need status updates at one or more monitors to be \emph{as timely as possible}; however, this is typically constrained by limited network resources. 

For example, consider 
various sensors (location, acceleration, tire pressure, etc.) in a vehicle generating status updates which the in-vehicle radio delivers to other vehicles in vicinity or other networked monitoring systems.
The update packets are queued while they wait to be serviced by the car radio. The packet currently being serviced by the radio waits for medium access and transmission before it is received by other cars. Note that each sensor in the car may be a source or  the car may aggregate a collection of sensor measurements into a status update message that is transmitted as a single packet.  
The packet service time will depend on the wireless channel and may or may not incorporate retransmissions due to channel errors and backoff due  to the activity of other wireless transmitters.  While system models that incorporate these effects can be arbitrarily complex,   we observe that optimal updating policies are not well understood even in the simple setting of  M/M/1 queues.

Maintaining the timeliness of data and state information in a network is a problem that has appeared in many forms,  including, for example,  data freshness in warehouses \cite{karakasidis_etl_2005} and web caches \cite{yu_scalable_1999}, periodic updating of real time databases \cite{xiong_deriving_1999}, and route caches in ad~hoc networks \cite{hu_ensuring_2002}. However, no consistent analytic methodology has emerged. This paper focuses on an {\em age of information (AoI)} timeliness metric as a basis for the evaluation and design of status update systems.  

When a monitor's most recently received update at time $t$ is timestamped $u(t)$,  the \emph{status update age} or simply the \emph{age}, is the random process $\age(t)=t-u(t)$ and  the AoI is the average $\age(t)$.  The monitor's requirement of timely updating corresponds to small AoI.  AoI is an application-independent metric that permits evaluation of the network performance, separate from application-specific metrics that may be too complex to employ in the design of the network. However, AoI can also be useful in specific applications by designing the communication network to meet statistical requirements, such as expected value and variance, of the age process.  For example, if a status updating system is reporting sample values of a Wiener process $X(t)$ with variance $\alpha t$ \cite{papoulis},  then the monitor's  MMSE estimate of $X(t)$ given the status age $\age(t)$ is $\hat{X}(t)=X(t-\age(t))$. The variance of this estimate is $\alpha\age(t)$.   

Traditionally, network performance has been characterized by tradeoffs in rate, delay, throughput and loss.  The data rate can be increased, but this induces additional delay in lossless systems or increased packet dropping in lossy systems. Furthermore, comparisons between lossless and lossy networks are generally problematic.  By contrast, we will see that AoI is fundamentally different; the age metric enables direct comparison of lossless and lossy systems. Moreover,  
the goal of timely updating is neither the same as maximizing the throughput or  utilization of the communication system, nor of ensuring that generated status updates are received with minimum delay.   Utilization is maximized when  sensors send updates as fast as possible. However, this can lead to the monitor receiving delayed updates because the status messages become backlogged in the communication system. Instead, we will see that  sources can minimize their AoI by optimizing their updating rates in response to the available system resources. 

We further observe that it may also be desirable to redesign systems to facilitate timely updating. 
A basic property of the first-come first-served (FCFS) queue model is that new update messages can be queued behind outdated messages that were generated earlier. This can be viewed as an undesirable consequence of protocol layering or of the hardware design.
However, among all status update packets in the wireless interface, the transmission of the youngest packet will minimize the status age at the monitor. Moreover, under the assumption that a status update carries the Markov state of the source, the transmission of the youngest status update obviates the need for transmission of the older outdated packets in the queue. Thus it is desirable to implement a lossy last-come first-served (LCFS) queueing discipline in which a new status update packet 
will preempt any previously queued update packets and this preempted packet will be discarded. 


\subsection{Prior Work and Related Applications}
\label{sec:related}
This paper expands on our analyses of status age in single-source single-server queues \cite{KaulYatesGruteser-Infocom2012}, the M/M/1 LCFS queue with preemption in service \cite{2012CISS-KaulYatesGruteser}, and the M/M/1 FCFS system with multiple sources \cite{2012ISIT-YatesKaul}. Other contributions to AoI analysis have also appeared recently.  To evaluate AoI for a single source sending updates through a network cloud \cite{KamKompellaEphremides2013ISIT} or through an M/M/2 server \cite{KamKompellaEphremides2014ISIT,Kam-PathDiversity2016}, out-of-order packet delivery was the key analytical challenge. A related (and generally more tractable) metric, peak age of information (PAoI), was introduced in \cite{CostaCodreanuEphremides2014ISIT}. Properties of PAoI have also been studied in \cite{HuangModiano2015ISIT}  for an FCFS M/G/1 multiclass queue. In  \cite{CostaCodreanuEphremides2014ISIT,CostaCodreanuEphremides2016}, the authors analyzed AoI and PAoI for M/M/1/1 and M/M/1/2 queues that discard arriving updates if the system is full and also for  a third queue, dubbed M/M/1/2*,  in which an arriving update would preempt a waiting update.  In this work, the M/M/1/2* queue is called the M/M/1 LCFS-W (Last Come First Served with preemption only in Waiting) queue and here we extend AoI results to a LCFS-W  system with multiple sources.

Most recently, optimality properties of a Last Generated First  Served (LGFS) service discipline when updates arrive out of order are identified in \cite{BedewySunShroff-ISIT2016}, packet deadlines are  found to improve AoI in \cite{KamKompellaNWE-ISIT2016}, AoI in the presence of errors is evaluated in \cite{ChenHuang-ISIT2016}, and LCFS 
with non-memoryless gamma-distributed service times is considered in \cite{NajmNasser-ISIT2016}. There have also been recent studies of energy-constrained updating \cite{Elif2015ITA,Yates2015ISIT,UpdateorWait-Infocom2016} in which updates 
are submitted to the server with knowledge of the server state.

In addition to these queue-theoretic AoI analyses,  the theme of ensuring ``freshness'' has also appeared in various application areas, including that of networks, real time databases and warehousing. 

In~\cite{kaul_minimizing_2011}, we look at minimizing the age of status updates sent by vehicles over a carrier-sense multiple access (CSMA) network. A local minimum for the  age can be approached using gradient descent; however, it is not known if this is a global minimum and is only seen to exist in simulations. In~\cite{sanjit_kaul_piggybacking_2011}, we show that allowing nodes to piggyback other nodes' status updates can lead to a smaller age. 

For safety-related intelligent transport system applications, an  \emph{Awareness Quality} metric \cite{kloiber2012update} captures how fine and up-to-date the application information is. The authors observe that default metrics like throughput and delay are unable to capture awareness and propose \emph{Update Delay}, which is the elapsed time between application updates.  In~\cite{6117108}, the authors propose to use an oldest packet drop mechanism instead of a tail drop policy to reduce the delay of the received information, via beacons, in vehicular networks.

In~\cite{karakasidis_etl_2005}, the authors want to maximize the freshness of data in warehouses to meet user demands. They estimate the queue length and delay at the warehouse staging area where updates wait before they are committed to the warehouse database. 
Experiments lead them to conclude that small queues are desirable. 

Web caching reduces the latency in returning a web page to a client. However, unless refreshed often enough, a cache will return stale web pages. The refresh rate is limited by the finite time it takes for a cache to be updated after the page has been updated at the server. In~\cite{yu_scalable_1999} the authors propose an architecture that limits the ``\emph{degree of staleness}'' of a cache. Our work, for fairly simple descriptions of the time it takes to update a cache, answers how often the cache must be refreshed such that its age is minimized.

%
In~\cite{xiong_deriving_1999} the authors look at periodic transactions updating real time databases. Each transaction updates the database with data that is associated with a deadline relative to when it is generated. In their work, there is no assumed limit on available processing power (service rate). The objective is to find the combination of update period and deadline such that all transactions complete before their deadlines, thus ensuring the freshness of data while minimizing the CPU utilization.

Ad hoc networking protocols typically use cached routes  to forward packets to their destinations. In~\cite{hu_ensuring_2002} the authors propose a mechanism that avoids propagation of stale route information. 
To avoid the overhead associated with periodic broadcasts of new route information, their method uses an epoch numbering system that helps network nodes to reject older information. In~\cite{giruka_hello_2005} the authors consider the issue of frequency of hello messages in ad-hoc networks. The frequency must not be so large as to congest the network but also not too small that the nodes have stale information.

Finally, data dissemination in sensor networks has been looked at under varied constraints. For example, in works like~\cite{he_energy-efficient_2004} and~\cite{schurgers_optimizing_2002} the authors consider energy efficient dissemination of state in sensor networks. More frequent updates lead to greater energy consumption and smaller sensor lifetime. Our work suggests strategies that a sensor, when awake, can use to minimize the average age of its status updates.

\subsection{Paper Overview} 


\begin{figure}[t]
\centering
\begin{tikzpicture}[node distance=1.1cm]
\node (vdots) {\raisebox{5pt}{$\vdots$}};
\node[draw,circle,thick] (s1)[above of = vdots] {\small\shortstack{Source\\ 1}};
\node[draw,circle,thick] (s2)[below of=vdots] {\small \shortstack{Source\\ $N$}};
\node (emid)[right of=vdots] {};
\node (enter)[right of=emid] {};
\node (smid)[right of=enter] {};
\draw [->, thick]  (s1.east) --  node [above right] {$\lambda_1$} (enter);
\draw [->, thick]  (s2.east) -- node [below right] {$\lambda_N$}(enter);
\node[draw,circle,thick] (server)[right of=smid] {$\mu$};
\draw[thick] (server.west) -- ++(0,0.5cm) --  ++(-1.75cm,0);
\draw[thick] (server.west) -- ++(0,-0.5cm) --  ++(-1.75cm,0);
\foreach \i in {1,...,3}
\draw[thick] (server.west) ++(-\i*0.5cm,0.5cm) -- ++(0,-1cm);
\node (monmid)[right of=server] {};
\node[draw,circle,thick] (mon)[right of=monmid] {\small Monitor};
\draw [->,thick] (server.east) -- (mon.west);
\end{tikzpicture}
\caption{Independent sources send status updates through a shared queue to a monitor.}
\label{fig:twosourcespic} 
\label{fig:model}
\vspace{-5mm}
\end{figure}
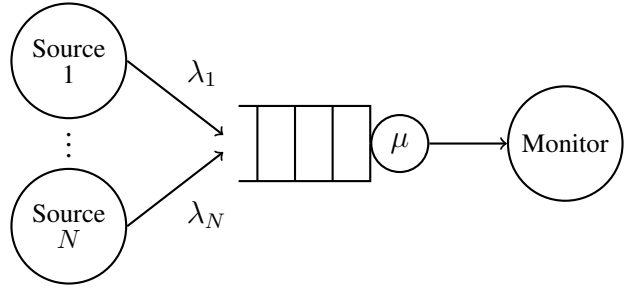
This work is based on the system depicted in Figure~\ref{fig:twosourcespic} in which a server delivers the updates of $N$ sources to a monitor.  Each source $i$ offers update packets as a rate $\lambda_i$ Poisson process. The service rate is $\mu$ for updates  from any source. This is sufficient to model systems in which the updating sources submit packets with identically distributed lengths but heterogeneous timeliness requirements.  Although the M/M/1 queue models that we examine are often too simple to describe practical networks,  status age is a new metric that is not well understood. We start here with these simple models to develop an understanding of the age of information in shared queues, in order to go on to characterize status age in more complex practical systems. 

In this system, source $i$ offers updating load $\rho_i=\lambda_i/\mu$ and the total offered load is 
\begin{equation}
\rho=\sum_{i=1}^N\rho_i. 
\end{equation}
The updates of source $i$ compete for the server against the aggregate other-source updating load 
\begin{equation}\eqnlabel{otherrho}
\OthersLoad{i}=\rho-\rho_i=\sum_{j\neq i} \rho_j.
\end{equation}

In Section~\ref{sec:ageprelim}, we  derive \Thmref{ageclaim}, a general result that describes the AoI $\age_i$ for each source $i$ in terms of the stationary properties of the interarrival times and system times of delivered source $i$  updates.
We then apply \Thmref{ageclaim} to M/M/1 systems  in which update packets arrive as  Poisson processes and have memoryless service times.  In particular, Section~\ref{sec:mm1FCFS} uses \Thmref{ageclaim} to derive the following result.  
\begin{theorem}\thmlabel{ageFCFS} $N$ sources with offered loads $\rho_1,\ldots,\rho_N$   and total load $\rho=\sum_{j}\rho_j$ at a rate $\mu$ M/M/1 FCFS queue  have average ages $\age_1,\ldots,\age_N$ such that
\begin{align*}
\age_i&=\frac{1}{\mu}\bracket{\frac{\rho_i^2(1-\rho\OthersLoad{i})}{(1-\rho)(1-\OthersLoad{i})^3}
+\frac{1}{1-\OthersLoad{i}}+\frac{1}{\rho_i}}.
\end{align*}
\end{theorem}

We continue the study of Poisson updaters with AoI results 
for a pair of lossy $N$-source M/M/1 LCFS systems. 
First, under LCFS \emph{with} preemption-in-service (denoted LCFS-S), a new update packet preempts any update packet currently in service. Second, under the LCFS with \emph{preemption only in waiting} (LCFS-W) queue discipline, a new packet replaces any older packet \emph{waiting} in the queue; however, the new packet has to \emph{wait} for any update packet currently in service to finish.  In this work, preemption is assumed to be source agnostic; we will allow a source's packet to be preempted by that of another source. Prioritized preemption policies are of considerable interest but beyond the scope of this work. For M/M/1 systems, the main result is summarized here:
\begin{theorem} \thmlabel{LCFS} $N$ sources with offered loads $\rho_1,\ldots,\rho_N$   at a rate $\mu$ M/M/1 LCFS queue with total load $\rho=\sum_{i=1}^N\rho_i$ have average ages $\age_1,\ldots \age_N$ such that
\begin{itemize}
\item[(a)] with preemption allowed in service (LCFS-S),
\begin{align*}
\age_i&=\frac{1}{\mu}(1+\rho)\frac{1}{\rho_i},
\end{align*}
\item[(b)] and with preemption allowed only in waiting (LCFS-W),
\begin{align*}
\age_i
&=\frac{1}{\mu}
\bracket{\alpha_W(\rho)+\paren{1+\frac{\rho^2}{1+\rho}}\frac{1}{\rho_i}}\\
\shortintertext{where}
\alpha_W(\rho)&=\frac{(1+\rho+\rho^2)^2+2\rho^3}{
(1+\rho+\rho^2)(1+\rho)^2}.
\end{align*}
\end{itemize}
\end{theorem}
We note that while $\alpha_W(\rho)$ is a ratio of fourth order polynomials, direct calculation will verify that
\begin{equation}\eqnlabel{alphaw-limits}
0.837 < \alpha_W(\rho)< 1.09,\qquad \rho\ge0.
\end{equation}

The proof of \Thmref{LCFS} appears in parts in various sections of this paper. In Section~\ref{sec:lcfsWithpre},
we use \Thmref{ageclaim} to derive \Thmref{LCFS}(a) for AoI in the LCFS-S queue. The method is similar to that used in   \cite{2012CISS-KaulYatesGruteser}, but with some algebraic simplifications that went previously unrecognized.
As it is based on \Thmref{ageclaim}, this analysis is conceptually similar to the FCFS analysis in Section~\ref{sec:mm1FCFS}. 

We note that for a single source with $\rho_1=\rho$, \Thmref{LCFS}(b) can be shown to reduce to the AoI of the M/M/1/2* queue, as given in \cite[Equation~(65)]{CostaCodreanuEphremides2016}.  We also note \Thmref{LCFS}(b) corrects an error in \cite[Equation~(23)]{2012CISS-KaulYatesGruteser}. In the context of a single-source system, this error was  identified and explained 
in \cite[Appendix]{CostaCodreanuEphremides2016}. That explanation serves to highlight how easily mistakes can be made in using the approach of \Thmref{ageclaim} for AoI analysis, even in simple memoryless-service systems. At the conclusion of \cite[Appendix]{CostaCodreanuEphremides2016}, the authors argue ``In the LCFS system with preemption we expect that, for very large arrival rates, the age would increase without bound, as no packet finishes service.'' We note that this speculation is contrary to the result of \Thmref{LCFS}(a), which in the special case of a single-source with $\rho_1=\rho=\lambda/\mu$, shows that the average age approaches $1/\mu$ as $\lambda\to\infty$.\footnote{For the single-source LCFS-S  system, this asymptotic result is a consequence of memoryless service. With fixed service rate $\mu$ and arrival rate $\lambda\to\infty$, the server is always occupied and (because the service is memoryless)  the queue departure rate approaches $\mu$. That is, the queue inter-departures approach a Poisson process of rate $\mu$. While the fraction $\mu/(\lambda+\mu)$ of those updates 
that complete service goes to zero, those that do complete service have system time $T$  that approaches zero. In this limiting case, the interarrival time $Y$ of a delivered update becomes an exponential $(\mu)$ random variable. In the  context of \Thmref{ageclaim} and the sawtooth age process in Figure~\ref{fig:age}, $\E{TY}\to0$, $\E{Y^2}\to2/\mu$ and $\E{Y}\to 1/\mu$.}

Nevertheless, the error in \cite{2012CISS-KaulYatesGruteser} and the skepticism expressed in  \cite{CostaCodreanuEphremides2016}  reflect on how difficult it can be to use \Thmref{ageclaim} to prove and verify AoI results, even for relatively simple service facilities. 
Thus we introduce in Section~\ref{sec:SHS} an analysis technique, namely stochastic hybrid systems (SHS) \cite{hespanha2006modelling}, that has not been previously applied to status updating systems. A stochastic hybrid system has a state with discrete components described by a Markov chain and continuous  components that are subject to reset mappings in discrete state transitions.  In AoI analysis, the queue state describing the number and source type of each update in the system is discrete while the age process at the monitor and the age of each update in the system varies continuously but is subject to reset mappings as updates enter or complete service, or get preempted.  

We will see that AoI tracking can be implemented as a simplified SHS with non-negative linear reset maps in which the continuous state is a piecewise linear process 
\cite{Vermes1980,GnedenkoKovalenko1966}, a special case of piecewise deterministic processes \cite{Davis1984,deville2016moment}. In this case, the SHS approach leads to a system of ordinary first order differential equations describing the temporal evolution of the expected value of the age process.  The SHS approach may not   appear to be simple at first; however,   in Section~\ref{sec:SHS-AoI}  it yields \Thmref{AOI-SHS}, a  simple, systematic (and largely mechanical) procedure for the calculation of AoI in finite-state queues with memoryless service.  

In Section~\ref{sec:LCFS-SHS}, we demonstrate the power of \Thmref{AOI-SHS} by using it to prove \Thmref{LCFS}. 
In particular, Section~\ref{sec:LCFS-SHS} exercises \Thmref{AOI-SHS} in a sequence of three SHS-based derivations of \Thmref{LCFS}(a). The first derivation is a straightforward application of \Thmref{AOI-SHS} with three discrete states to track whether the system is idle or busy, and whether an update in service is from the source of interest. The second derivation demonstrates how the discrete state space can be reduced to two states (idle or busy) by careful embedding of some elements of the discrete state in the continuous state. The third derivation shows how \emph{fake updates} can be used to reduce the LCFS-S system to a single discrete state. The embedding approach is then used in Section~\ref{sec:lcfsWOpreMM1} to provide an SHS derivation of \Thmref{LCFS}(b) for AoI in the LCFS-W system.  We note that SHS analysis of the LCFS-S and LCFS-W systems is far simpler than analyses based on \Thmref{ageclaim}. In Section~\ref{sec:SHS-matrix}, we rewrite the equations of \Thmref{AOI-SHS} in a non-negative matrix form in order to prove \Thmref{AOI-SHS}.

Finally, in Section~\ref{sec:perf-eval} we return to  examine the performance of the updating system shown in Figure~\ref{fig:twosourcespic}. We use Theorems~\thmref{ageFCFS} and \thmref{LCFS} to examine achievable AoI regions for two-source FCFS and LCFS systems. In addition, resource sharing issues for $N$ sources  are also explored. Our results show that there are nontrivial gains in trunking efficiency when $N$ sources share the system capacity with coordinated load balancing of the sources. In particular, high offered load at an FCFS system induces high AoI through queueing delays. A lossy LCFS discipline can mitigate this problem but its packet discarding policy may encourage sources to operate at excessively high offered loads.
A short conclusion follows in Section~\ref{sec:conclusions}.


%
%
%

\subsection{Notation}
For integers $m\le n$, $\range[m]{n}=\set{m,m+1,\ldots,n}$; otherwise $\range[m]{n}$ is an empty set. 
The vectors $\zerov[n]$ and $\onev$ denote the row vectors $\rowvec{0 & 0 &\cdots& 0}$ and  $[\begin{matrix}1 & 1&\cdots 1\end{matrix}]$ in $\R^n$. A vector $\xv\in\R^n$ is  a $1\times n$ row vector with elements $\rowvec{x_0&x_1&\dots&x_{n-1}}$.   Similarly, a matrix $\Bmat\in\R^{n\times n}$ has elements $[\Bmat]_{i,j}$ for $i,j\in 0:(n-1)$.
For  a vector $\xv$ and a matrix $\Bmat$, $[\xv]_j$ and $[\Bmat]_j$ denote the $j$th element and $j$th column respectively for $j=0,\ldots,n-1$. For a vector process $\xv(t)$, we use $\dot{\xv}$ and $\dot{\xv}(t)$  to denote the derivative $d\xv(t)/dt$.
\section{Time-Average Age Analysis}
\label{sec:ageprelim}

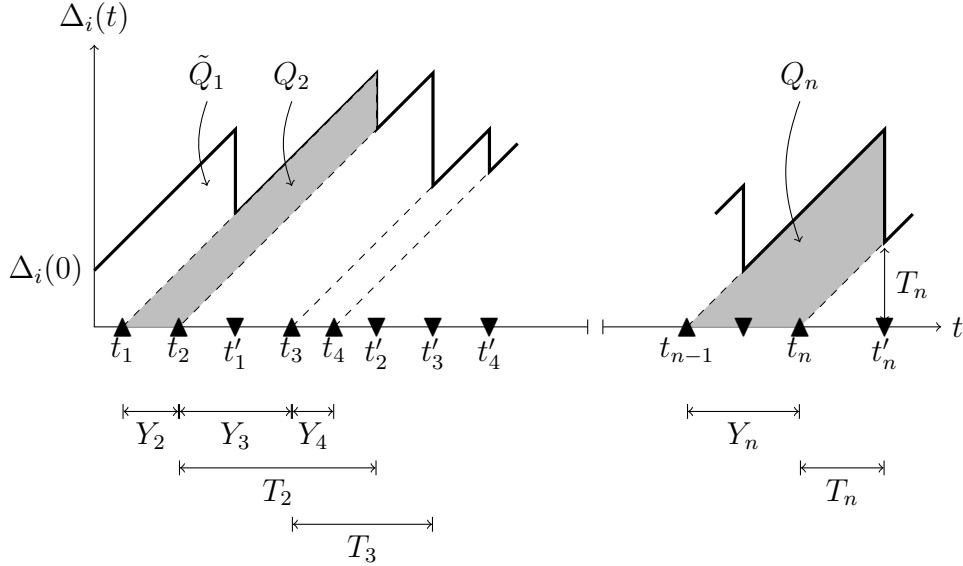
\begin{figure}[t]
\centering
\begin{tikzpicture}[scale=\linewidth/44cm]
\draw [<-|] (0,10) node [above] {$\age_i(t)$} -- (0,0) -- (17.5,0);
\draw [|->] (18,0) -- (30,0) node [right] {$t$};
\draw [very thick] (0,2) -- (5,7) -- (5,4)  -- (10,9) 
-- (10,7)  -- (12,9) -- (12,5) -- (14,7) -- (14,5.5) -- (15,6.5);
\draw [fill=lightgray, ultra thin, dashed] (1,0) to (10,9) to (10,7) to (3,0);
\draw 
(0,2) node [left] {$\age_i(0)$}
(1,0) node {$\blacktriangle$}
(1,0) node [below] {$t_1$} 
(3,0) node {$\blacktriangle$}
(3,0) node [below] {$t_2$} 
(5,0) node {$\blacktriangledown$}
(5,0) node [below] {$t'_1$}
(7,0) node {$\blacktriangle$}
(7,0) node [below] {$t_3$}
(8.5,0) node {$\blacktriangle$}
(8.5,0) node [below] {$t_4$}
(10,0) node {$\blacktriangledown$}
(10,0) node [below] {$t'_2$}
(12,0) node {$\blacktriangledown$}
(12,0) node [below] {$t'_3$}
(14,0) node {$\blacktriangledown$}
(14,0) node [below] {$t'_4$};
\draw[<-] (4,5) to [out=110,in=250] (4,8) node [above] {$\tilde{Q}_1$};
\draw[<-] (7,5) to [out=110,in=250] (7,8) node [above] {$Q_2$};
\draw  [|<->|] (1,-3) to node [below] {$Y_2$} (3,-3);
\draw  [|<->|] (3,-3) to node [below] {$Y_3$} (7,-3);
\draw  [|<->|] (7,-3) to node [below] {$Y_4$} (8.5,-3);
\draw  [|<->|] (3,-5) to node [below] {$T_2$} (10,-5);
\draw  [|<->|] (7,-7) to node [below] {$T_3$} (12,-7);
\draw [thin, dashed] (7,0) to (12,5);
\draw [thin, dashed] (8.5,0) to (14,5.5);
\draw [fill=lightgray, ultra thin, dashed] (21,0) to (28,7) to (28,3) to (25,0);
\draw  
(21,0) node {$\blacktriangle$}
(21,0) node [below] {$t_{n-1}$}
(23,0) node  {$\blacktriangledown$}
(25,0) node {$\blacktriangle$}
(25,0) node [below] {$t_{n}$}
(28,0) node  {$\blacktriangledown$}
(28,0) node [below] {$t'_n$};
\draw [very thick] (22,4) -- (23,5) -- (23,2) -- (28,7) -- (28,3) -- (29,4); 
\draw  [|<->|] (21,-3) -- node [below] {$Y_n$} (25,-3);
\draw  [|<->|] (25,-5) -- node [below] {$T_{n}$} (28,-5);
\draw [<->] (28,2.8) -- node [right] {$T_n$} (28,0.2); 
\draw[<-] (25,3) to [out=110,in=250] (25,8) node [above] {$Q_{n}$};
\end{tikzpicture}
\caption{\small Example change in status update age at a monitor for a system with a FCFS queue. Updates from source $1$ arrive at times marked $\blacktriangle$ and are received at the monitor at times marked $\blacktriangledown$.}
\label{fig:age}
\vspace{-5mm}
\end{figure}

Figure~\ref{fig:age} shows a sample variation of age $\age_i(t)$, for source $i$ as a function of time $t$, at the monitor. Without loss of generality, assume that we begin observing at $t=0$ when the queue is empty and the age is $\age_i(0)$. 
The first status update of source $i$ is timestamped $t_1$ and is followed by updates timestamped $t_2,t_3,\ldots,t_n$.
The status age of source $i$ at the monitor increases linearly in time in the absence of any updates and is reset to a smaller value when an update is received. 
Update $j$ of source $i$, generated at time $t_j$, finishes service and is received by the monitor at time $t'_j$. At $t'_j$, the age $\age_i(t'_j)$ at the monitor is reset to the age $T_j = t'_j - t_j$ of the received status update. The age $T_j$ is also the system time of update packet $j$.
Thus the age function $\age_i(t)$ exhibits the sawtooth pattern shown in Figure~\ref{fig:age}. 
The time average age of the status updates is the area under the age graph in Figure~\ref{fig:age} normalized by the time interval of observation.%

Over an interval $(0,\mathcal{T})$, the average age is
\begin{align}
\Tavg{\age_i} = \frac{1}{\mathcal{T}}\int_{0}^{\Tcal} \age_i(t) dt.
\label{eqn:averageAge}
\end{align}
For simplicity of exposition,  the length of the observation interval is chosen to be $\mathcal{T}=t'_n$, as depicted in  Figure~\ref{fig:age}. We decompose the area defined by the integral \eqnref{averageAge} into the  sum of 
the polygon area $\tilde{Q}_1$, the trapezoidal areas $Q_j$ for $j\ge 2$ ($Q_2$ and $Q_n$ are highlighted in the figure), and the triangular area of width $T_n$ over the time interval $(t_n,t_n')$. 
From Figure~\ref{fig:age}, we see that $Q_j$ can be calculated as the difference between the area of the isosceles triangle whose base connects the points $t_{j-1}$ and $t'_j$ and the area of the isosceles triangle with base connecting the points $t_j$ and $t'_{j}$. 
Defining 
\begin{equation}
Y_j=t_j - t_{j-1}
\end{equation}
to be the interarrival time of update $j$, it follows that
\begin{align}\eqnlabel{Qi}
Q_j = \frac{1}{2}(T_{j} + Y_j)^2 - \frac{1}{2}T_{j}^2 = Y_j T_{j}+Y_j^2/2.
\end{align}
With  $N_i(\mathcal{T}) = \max\{n|t_n\le \mathcal{T}\}$ denoting the number of source $i$ updates by time $\mathcal{T}$, this decomposition,
%
%
 along with some rearrangement, yields the time-average age 
\begin{align}
\Tavg{\age_i} &= \frac{\tilde{Q}}{\mathcal{T}}
+ \frac{(N_i(\mathcal{T})-1)}{\mathcal{T}}
\frac{\sum_{j=2}^{N_i(\mathcal{T})} 
Q_j}{N_i(\mathcal{T})-1}
\eqnlabel{delta_tau_2}
\end{align}
where $\tilde{Q} = \Qtil_1+T_n^2/2$. We observe that the age contribution $\tilde{Q}$ represents a boundary effect that is finite with probability 1, so  the first term in~\eqnref{delta_tau_2} will vanish as $\Tcal$ grows. 

\begin{definition}\label{def:ergodic-update}
A status updating system for a source $i$ is stationary and ergodic if  $(Y_j,T_j)$ is a stationary sequence with marginal distribution identical to $(Y,T)$, and as $\mathcal{T}\to\infty$, 
\begin{align*}
 \frac{N_i(\mathcal{T})}{\mathcal{T}}\to\frac{1}{\E{Y}}, \quad\text{and}\quad
\frac{\sum_{j=2}^{N_i(\mathcal{T})}
Q_j}{N_i(\mathcal{T})-1} \to\E{Q}
\end{align*}
with probability $1$.
\end{definition}
For such systems, the AoI of source $i$ is 
$\age_i=\limty{\mathcal{T}}\Tavg{\age_i}$ 
and \eqnref{delta_tau_2} implies the next claim. 
\begin{theorem}\thmlabel{ageclaim}
For a stationary  ergodic status updating system in which $Y$ is the interarrival time between delivered source $i$ updates and $T$ is the system time of such a delivered packet, the AoI for source $i$ is 
\begin{align*}
\age_i&=\frac{\E{Q}}{\E{Y}}=\frac{\E{YT} +\E{Y^2}/2}{\E{Y}}.
\end{align*}
\end{theorem}

We note that \Thmref{ageclaim} is more akin to a bookkeeping identity such as Little's Law in that sufficient conditions for the ergodicity of the age process are not explicitly provided
but can be verified for most reasonably-designed   service systems. 
As a consequence, \Thmref{ageclaim} can be applied to a broad class of  service systems, including both lossless FCFS systems as well as lossy LCFS systems in which updates are preempted and discarded.  Furthermore, it makes no specific assumptions regarding other traffic that might share the queue with the update packets of source $i$.

With respect to \Thmref{ageclaim}, we emphasize that $Y$ is the interarrival time between \emph{delivered} updates of source $i$, and $T$ is  the system time of a \emph{delivered} update. These  somewhat cumbersome definitions are a consequence of the generality of the approach.  For example, Section~\ref{sec:mm1FCFS} employs \Thmref{ageclaim} to evaluate a work-conserving M/M/1 FCFS system in which the $Y_j$ are independent identically distributed (iid) exponential interarrival times and the $T_j$ are the  packet system times. 
On the other hand,  Section~\ref{sec:lcfsWithpre} uses \Thmref{ageclaim} to analyze the LCFS-S system that supports preemption of the packet in service. In this system, packet $j$ refers to the $j$th packet that completes service and is delivered to the monitor. 
There may be an arbitrarily large number of update packets that arrive between packets $j-1$ and $j$ that are preempted and discarded before completing service. 


\subsection{M/M/1 First-Come First-Served}
\label{sec:mm1FCFS}

In prior work \cite{KaulYatesGruteser-Infocom2012}, we analyzed M/M/1 FCFS queues serving the status updates of a single source.  
In that work, it was shown that the average status age for an M/M/1 queue with arrival rate $\lambda$, service rate $\mu$ and offered load $\rho=\lambda/\mu$ is given by 
\begin{align}
\age = \frac{1}{\mu}\bracket{\frac{\rho^2}{1-\rho}+1+\frac{1}{\rho}}.
\eqnlabel{sysAgeMM1fcfs}
\end{align}
The average age $\age$ in \eqnref{sysAgeMM1fcfs} is minimized at $\rho^*\approx 0.53$. In this section, we generalize this result to an $N$ source system.

Following \Thmref{ageclaim}, the system time of a source $i$ update packet 
is
$T=W+S$,
where $W$ and $S$ are the respective waiting and service times.
Since 
$S$ is independent of $Y$, it follows 
that
$\E{YT}
= \E{YW}+ \E{Y}\E{S}$.
We note that $\E{S} = 1/\mu$ and that the rate $\lambda_i$ Poisson arrival process implies $\E{Y}= 1/\lambda_i$.
and $\E{Y^2}=2/\lambda_i^2$. 
It follows from \Thmref{ageclaim}
that
\begin{equation}
\age_i=\lambda_i\E{YW}+\frac{1}{\mu}+\frac{1}{\lambda_i}.\eqnlabel{age-WX2}
\end{equation}
The expectation $\E{YW}$ is nontrivial because $Y$ and $W$ are negatively correlated; a large interarrival time $Y$ can allow the queue to empty, yielding a small waiting time $W$. 
Evaluation of $\E{YW}$ is provided in Appendix~\ref{sec:EWXproof} in the proof of the following lemma. 
 \begin{lemma}\label{lem:EWX}
\begin{equation*}
\E{YW} = \frac{1}{\mu^2}\bracket{\frac{\rho_i(1-\rho\OthersLoad{i})}{(1-\rho)(1-\OthersLoad{i})^3}
+\frac{\OthersLoad{i}}{\rho_i(1-\OthersLoad{i})}}.
\end{equation*}
\end{lemma}

Applying Lemma~\ref{lem:EWX} to \eqnref{age-WX2} yields \Thmref{ageFCFS}. We note that \Thmref{ageFCFS} reduces to the single source result \eqnref{sysAgeMM1fcfs} when $\rho_i=\rho$ and  $\OthersLoad{i}=0$. 


\subsection{LCFS With Preemption In Service: Analysis}
\label{sec:lcfsWithpre}
In this system, a packet arrival preempts the packet currently in service, if any. 
The number of packets in such a system is at most $1$. To analyze this system, we start with \Thmref{ageclaim}. 
As shown in Figure~\ref{fig:age}, 
update packets generated by source $i$ at time instants $t_j$
are those updates that \emph{complete} service and  $Y_j=t_j-t_{j-1}$ is the time between such arrivals. The service time (and also system time) of this $j$th packet is $\Spre_j$.

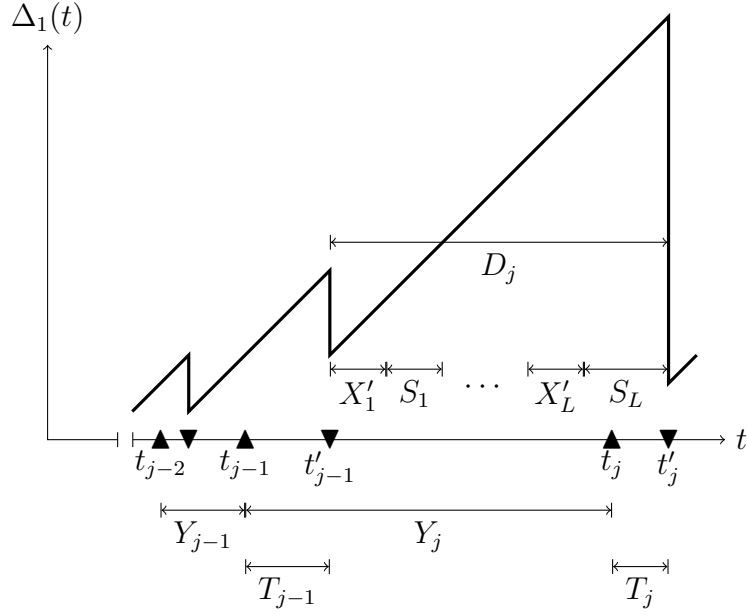
\begin{figure}[t]
\centering
\begin{tikzpicture}[scale=\linewidth/44cm]
\draw [<-|] (0,14) node [above] {$\age_1(t)$} -- (0,0) -- (2.5,0);
\draw [|->] (3,0) -- (24,0) node [right] {$t$};
\draw [very thick] (3,1) -- (5,3) -- (5,1)  -- (10,6) -- (10,3)  -- (22,15) 
-- (22,2) -- (23,3);
\draw 
(4,0) node {$\blacktriangle$}
(4,0) node [below] {$t_{j-2}$} 
(5,0) node {$\blacktriangledown$}
(7,0) node {$\blacktriangle$}
(7,0) node [below] {$t_{j-1}$} 
(10,0) node {$\blacktriangledown$}
(10,0) node [below] {$t'_{j-1}$}
(20,0) node {$\blacktriangle$}
(20,0) node [below] {$t_j$}
(22,0) node {$\blacktriangledown$}
(22,0) node [below] {$t'_j$};
\draw  [|<->|] (4,-2.5) to node [below] {$Y_{j-1}$} (7,-2.5);
\draw  [|<->|] (7,-2.5) to node [below] {$Y_j$} (20,-2.5);
\draw  [|<->|] (7,-4.5) to node [below] {$\Spre_{j-1}$} (10,-4.5);
\draw  [|<->|] (20,-4.5) to node [below] {$\Spre_j$} (22,-4.5);
\draw  [|<->|] (10,2.5) to node [below] {$X'_1$} (12,2.5);
\draw  [|<->|] (12,2.5) to node [below] {$S_1$} (14,2.5);
\draw (15.5,2.5) node [below] {$\cdots$};
\draw  [|<->|] (17,2.5) to node [below] {$X'_L$} (19,2.5);
\draw  [|<->|] (19,2.5) to node [below] {$S_L$} (22,2.5);
\draw  [|<->|] (10,7) to node [below] {$D_j$} (22,7);
\end{tikzpicture}
\caption{\small Example change in update age of source $i$ under LCFS \emph{with} preemption in service. On the time axis, updates from source $i$ arrive at times marked $\blacktriangle$ and are received at the monitor at times marked $\blacktriangledown$.}
	\label{fig:age_lcfs_with_preempt}
\end{figure}

In order to calculate the average age $\age_i$, let $\interDep_j$ (see  Figure~\ref{fig:age_lcfs_with_preempt}) be the time interval between the departures $j-1$ and $j$. This interval starts with an idle period and may see zero or more arrivals of other sources, some of which may complete service, while others are preempted. Any arrivals of the given source during $\interDep_j$, other than arrival $j$, are preempted. Thus the interval $\interDep_j$ consists of one or more blocks of the  server being idle followed by it being busy. Note that if the system consists of just one source, then $\interDep_j$ consists of just one block, which starts with the idle period that follows the departure of $j-1$. This idle period is followed by the server busy period that ends in departure $j$. Figure~\ref{fig:age_lcfs_with_preempt} shows $\interDep_j$, which contains a random $L$ number of blocks. The figure shows blocks $1$ and $L$. A block $k$, say of length $\interDblock_k$, consists of an idle period of length $X'_k$ followed by a busy period of length $S_k$. We have
\begin{align}
\interDep_j= \sum_{k=1}^{L} \interDblock_k = \sum_{k=1}^{L} (X'_k + S_k).
\label{eqn:2userB}
\end{align}
Note that packet $j$ arrives during $S_L$ and then spends time $\Spre_j$ in service.

We will now calculate the terms $\E{Y}$, $\E{Y^2}$ and $\E{YT}$ in \Thmref{ageclaim} in terms of $\interDep_j$ and $\Spre_j$. Consider the interval $Y_j$, for any $j$. We observe from Figure~\ref{fig:age_lcfs_with_preempt} that
\begin{align}
Y_j +\Spre_j = \Spre_{j-1} + \interDep_{j} \eqnlabel{y3}.
\end{align}
Because $\Spre =^{st} \Spre_{j-1} =^{st} \Spre_{j}$, $Y =^{st} Y_j$, and $\interDep =^{st} \interDep_j$,
\begin{align}
\E{Y} = \E{Y_j} = \E{\interDep_{j}} = \E{\interDep}.
\eqnlabel{e_y_3}
\end{align}
Note that $Y_j$ and $\Spre_j$ are independent. Thus \eqnref{e_y_3} implies 
\begin{align}
\E{Y_j\Spre_j} = \E{Y_j}\E{\Spre_j} = \E{Y}\E{\Spre} = \E{\interDep}\E{\Spre}.
\eqnlabel{e_y_3_s_3}
\end{align}
Furthermore, since 
$\interDep_j$ and $\Spre_{j-1}$ are also independent, it also follows from  \eqnref{y3} that
\begin{equation}
\Var{Y_j} +\Var{\Spre_j} = \Var{\Spre_{j-1}} + \Var{\interDep_{j}}.
\end{equation}
It then follows from \eqnref{e_y_3} that
$\E{Y^2} = \E{\interDep^2}$.
This fact, combined with (\ref{eqn:e_y_3}) 
and~(\ref{eqn:e_y_3_s_3}), simplify \Thmref{ageclaim} 
to 
\begin{align}
\age_i = \E{\Spre} + \frac{\E{\interDep^2}}{2\E{\interDep}}.
\eqnlabel{age_lcfs_with_p}
\end{align}
The remainder of the proof of \Thmref{LCFS}(a), specifically the calculation of the moments  in \eqnref{age_lcfs_with_p}, appears in  Appendix~\ref{sec:Proof:LCFS-S}.





\section{Stochastic Hybrid Systems for AoI Analysis}\label{sec:SHS}
We start in Section~\ref{sec:SHSintro} with an introduction to the key elements in the general SHS method. 
In Section~\ref{sect:SHSpiecewise}, we consider a special case of SHS in which the continuous state $\xv(t)$ is a piecewise linear process that is subject to linear reset maps during a discrete state transition. For the piecewise linear SHS, we derive a set of first order differential equations for the first order moments of the continuous state. Section~\ref{sec:SHS-AoI} employs the piecewise linear SHS to derive  the AoI of a general finite state queueing system described by a continuous-time Markov chain. The resulting methodology makes AoI computation practically as simple as the calculation of stationary probabilities of the Markov chain of the queue. 

\subsection{A Brief Introduction to SHS}\label{sec:SHSintro}
There are many SHS variations \cite{teel2014stability}, but in this work we follow the model and notation in \cite{hespanha2006modelling}. In an SHS, the state is partitioned into a discrete component  $q(t)\in\Qcal=\set{0,1,\ldots,m}$ that evolves as a point process and a continuous component  $\xv(t)=\rowvec{x_0(t) & \cdots &x_n(t)}\in\R^{n+1}$.  Given the discrete set  $\Qcal$ and the $k$-vector $\zv(t)$ of independent Brownian motion processes,  an SHS is defined by a stochastic differential equation  
 \begin{equation}\eqnlabel{SHSde}
 \dot{\xv}=f(q,\xv,t) + g(q,\xv,t)\dot{\zv}
 \end{equation}
 for mappings $f:\Qcal\times \R^{n+1}\times[0,\infty)\to\R^{n+1}$ and $g: \Qcal\times \R^{n+1}\times [0,\infty)\to\R^{(n+1)\times k}$, and a set of transitions $\Lcal=\set{1,\ldots,\ell_0-1}$, such that each $l\in \Lcal$ defines  a discrete transition/reset map 
 \begin{subequations}
 \eqnlabel{SHStrans}
 \begin{align}
 (q',\xv') &=\phi_l(q,\xv,t), & \phi_l:\Qcal\times \R^{n+1}\times [0,\infty)&\to\Qcal\times\R^{n+1},\eqnlabel{SHSjump}\\
 \shortintertext{with transition intensity}
 &\laml(q,\xv,t), & \laml:\Qcal\times\R^{n+1}\times [0,\infty)&\to [0,\infty).
 \end{align}
 \end{subequations}
When the system is in discrete state $q$, $\xv(t)$ evolves according to \eqnref{SHSde}; but  in a discrete transition from $q$ to $q'$, the continuous state can have a discontinuous jump from $\xv$ to $\xv'$, as described by \eqnref{SHSjump}. The resulting $\xv(t)$ process has piecewise continuous sample paths. Associated with each transition $l$ is a counting process $N_l(t)$ that counts the number of occurrences of transition $l$ in the interval $[0,t]$. The probability that $N_l$ jumps in the interval $(t,t+dt]$ is $\laml(q(t),\xv(t),t)\,dt$.     

Because of the generality and power of the SHS model, characterization of the $q(t)$ and $\xv(t)$ processes can be complicated and often intractable. The approach in \cite{hespanha2006modelling} is to define test functions $\psi(q,\xv,t)$ whose expected values $\E{\psi(q(t),\xv(t),t)}$ can be evaluated as functions of time. %
We refer the reader to \cite{hespanha2006modelling} and the survey \cite{teel2014stability} for additional background.
\subsection{Piecewise linear SHS with linear reset maps}
\label{sect:SHSpiecewise}
In the setting of status updates passing through queues with memoryless service processes, we restrict our attention to systems in which  $q(t)$ is a continuous-time finite-state Markov chain that describes the occupancy of a service facility and $\xv(t)\in\R^{n+1}$ describes the continuous-time evolution of a collection of  age-related processes. 

In particular, we consider a restricted class of SHS in which the components of $\xv(t)$ are deterministic constant-slope ramp processes that can have discontinuous jumps during discrete state transitions. We will see that this will be sufficient to capture the sawtooth age processes. In terms of the general SHS model given by \eqnref{SHSde} and \eqnref{SHStrans}, we have
\begin{subequations}\eqnlabel{simplified}
\begin{align}
f(q,\xv,t)&=\bv_q,\eqnlabel{simplified-f}\\
g(q,\xv,t)&=0,\eqnlabel{simplified-g}\\
\laml(q,\xv,t)&=\laml\dq{\ql},\eqnlabel{simplified-lam}\\
\phi_l(q,\xv,t) &=(q'_l,\xv\Amat_l).\eqnlabel{simplified-map}
\end{align}
\end{subequations}
In the graphical representation of the Markov chain $q(t)$, each state $q\in\Qcal$ is a node and each transition $l$ is a directed edge $(q_l,q'_l)$ with transition rate $\laml$ while $q(t) = q_l$. 
In \eqnref{simplified-lam}, the Kronecker delta function $\dq{q_l}$ ensures that transition $l$ occurs only in state $q_l$. For each transition $l$, the transition reset map will be a linear mapping of the continuous state $\xv$ of the form $\xv'=\xv\Amat_l$.  That is, transition $l$ causes the system to jump to discrete state $q'_l$ and resets the continuous state from $\xv$ to $\xv'=\xv\Amat_l$.   In addition, we note that \eqnref{simplified-f} and \eqnref{simplified-g} imply that the continuous state evolution \eqnref{SHSde} in each discrete state $q(t) = q$ is  
\begin{align}\eqnlabel{xv-evolution}
\dot{\xv}(t)=\bv_q.
\end{align}
Thus, the evolution of $\xv(t)$ in each state is specified by $\Bcal=\set{\bv_q:q\in\Qcal}$. Furthermore, the transition links $l$ are described by  the tuples $a_l=(q_l,q'_l,\laml,\Amat_l)$ and the set of transitions is $\Acal=\set{a_l: l\in\Lcal}$.  Thus   we refer to a piecewise linear SHS with linear reset maps by the tuple  $(\Qcal,\Bcal,\Acal)$.

The transition rates $\set{\laml}$ correspond to the transition rates associated with the continuous-time Markov chain for the discrete state $q(t)$; but there are some differences. Unlike an ordinary continuous-time Markov chain, the SHS may include self-transitions in which the discrete state is unchanged because a reset occurs in the continuous state. Furthermore, for a given pair of states $i,j\in\Qcal$, there may be multiple transitions $l$ and $l'$ in which the discrete state jumps from $i$ to $j$ but the transition maps $\Amat_l$ and $\Amat_{l'}$ are different.\footnote{For example, consider a queueing system in which an update in service can either complete service or be discarded in the middle of service. Under either transition, the next discrete state reflects the departure or discard of the update in service. However, a service completion yields a reduction in age while discarding an update in service results in no reduction in age.}

For each $\qhat\in\Qcal$,
it will be sufficient for average age analysis to  employ test functions of the form $\psi(q,\xv)=\psim[]{\qhat}(q,\xv)$ and $\psi(q,\xv)=\psim{\qhat}(q,\xv)$ such that 
\begin{subequations}\eqnlabel{testfcn-S:defn}
\begin{align}
\psim[]{\qhat}(q,\xv) &= \dq{\qhat}\\ 
\shortintertext{and}
\psim{\qhat}(q,\xv) &= x_j\dq{\qhat},\qquad j\in\range{n}.
\end{align}
\end{subequations}
Based on these test functions, we define for all $\qhat\in\Qcal$,
\begin{subequations}\eqnlabel{pivi-defns}
\begin{align}
\pi_{\qhat}(t) &= \E{\psim[]{\qhat}(q(t),\xv(t))}=\E{\dqt{\qhat}},
\eqnlabel{piqhat-defn}\\
v_{\qhat j}(t)&=\E{\psim{\qhat}(q(t),\xv(t))}=\E{x_j(t)\dqt{\qhat}},\quad  j\in\range{n},\eqnlabel{vqi-defn}\\
\shortintertext{and the vector functions}
 \vv_{\qhat}(t)&=\cvec{v_{\qhat 0}(t),\dots,v_{\qhat n}(t)}=\E{\xv(t)\dqt{\qhat}}. \eqnlabel{vv-defn}
 \end{align}
\end{subequations}

We note that $\pi_{\qhat}(t)$ denotes the discrete Markov state probabilities, i.e.,
\begin{equation}
\pi_{\qhat}(t)=\E{\dqt{\qhat}}=\prob{q(t)=\qhat}.
\end{equation} 
Similarly, $\vv_{\qhat}(t)$ measures correlation between the age process $\xv(t)$ and the discrete state $q(t)$.

Associated with an SHS
is a mapping $\psi\to L\psi$ known as the extended generator. From \cite[Theorem~1]{hespanha2006modelling}, it follows from the conditions \eqnref{simplified} and the time invariance of $\psi(q,\xv)$ in \eqnref{testfcn-S:defn} that 
the extended generator of a piecewise linear SHS is given by
\begin{align}\eqnlabel{Lpsi-defn}
(L\psi)(q,\xv)
&= \frac{\partial\psi(q,\xv)}{\partial \xv}\cdot\bv_q
\TWOCOLUMN{\nn &\quad}+\sum_{l\in\Lcal} \paren{\psi(\phi_l(q,x))-\psi(q,x)}\laml(q),
\end{align}
where $\partial\psi(q,\xv)/\partial\xv$ denotes the gradient. 
Each test function $\psi(q(t),\xv(t))$  must satisfy Dynkin's formula
\begin{align}
\eqnlabel{dynkins}
\deriv{}{\E{\psi(q(t),\xv(t))}}{t}&=\E{(L\psi)(q(t),\xv(t))}.
\end{align}
Defining   
\begin{subequations}\eqnlabel{Lcalqbar}
\begin{align}
\Lcal'_{\qbar}&=\set{l\in\Lcal: q'_l=\qbar},\\
\Lcal_{\qbar}&=\set{l\in\Lcal: q_l=\qbar}
\end{align}
\end{subequations}
as the respective sets of incoming and outgoing  transitions for each state $\qbar$, we can now prove that $\pi_{\qbar}(t)$ and $\vv_{\qbar}(t)$ obey the 
system of first order ordinary differential equations given in the following lemma.

\begin{lemma}\label{lem:pi-vv-derivs}  For a piecewise linear SHS with linear reset maps $(\Qcal,\Bcal,\Acal)$,
\begin{subequations}
\begin{align}
\dot{\pi}_{\qbar}(t)&=\sum_{l\in\Lcal'_{\qbar}}\laml
\pi_{\ql}(t)- \pi_{\qbar}(t)\sum_{l\in\Lcal_{\qbar}}\laml,\qquad \qbar\in\Qcal,
\eqnlabel{pi-derivs}\\
\dot{\vv}_{\qbar}(t)&=
\bv_{\qbar}\pi_{\qbar}(t)+\sum_{l\in\Lcal'_{\qbar}}\laml \vv_{\ql}(t)\Amat_l
-\vv_{\qbar}(t)\sum_{l\in\Lcal_{\qbar}}\laml,\quad \qbar\in\Qcal.\eqnlabel{vv-derivs}
\end{align}
\end{subequations}
\end{lemma}
%
Proof of  Lemma~\ref{lem:pi-vv-derivs} appears in Appendix~\ref{sec:lem:pi-vv-derivs}.
From a given initial condition at time $t=0$, we can use Lemma~\ref{lem:pi-vv-derivs} to compute the temporal evolution of the discrete state probabilities $\pi_{\qbar}(t)$ and the expected values
$\vv_{\qbar}(t)=\E{\xv(t)\dqt{\qbar}}$.  Moreover, since
\begin{align}\eqnlabel{x-decomposition}
\xv(t)=\sum_{\qbar\in\Qcal} \xv(t) \delta_{\qbar,q(t)},
\end{align} 
Lemma~\ref{lem:pi-vv-derivs} enables us to compute  the expected value 
\begin{align}
\E{\xv(t)}=\sum_{\qbar\in\Qcal} \E{\xv(t)\delta_{\qbar,q(t)}}
=\sum_{\qbar\in\Qcal} \vv_{\qbar}(t).
\end{align}

\subsection{An SHS for AoI}\label{sec:SHS-AoI} 
We now employ a piecewise SHS with linear reset maps $(\Qcal,\Bcal,\Acal)$ for age tracking in a system described by  a continuous-time Markov chain. Our approach is to label the source of interest as source $1$ and to employ the continuous state $\xv(t)$ as a vector of age-related processes that enables tracking of the age of source $1$ updates at the monitor.

In the LCFS-S and LCFS-W systems with $\xv(t) = \rvec{x_0(t) & \ldots & x_n(t)}$  that we examine in Section~\ref{sec:LCFS-SHS}, $x_0(t)$ is the age at the monitor, $n$ is the maximum number of updates in the system, and updates in the system are indexed $1,2,\ldots,n$ such that if update $i$ is from source $1$, then $x_{i}(t)$, $1\le i\le n$, specifies the age of the update. In such systems, we will calculate the AoI $\age=\limty{t}\E{x_0(t)}$ of the source $1$ update process. The sequence of SHS analyses of the LCFS-S system will reveal there is considerable flexibility in choosing the continuous state $\xv(t)$. Moreover, careful definition of $\xv(t)$ can reduce the size of the discrete state space.

For example, in SHS analyses of the LCFS-S queue in Section~\ref{sec:LCFS-SHS}, $\xv(t)= [x_0(t),x_1(t)]$, where $x_0(t)$ tracks the age (of source~$1$ updates) and $x_1(t)$ is a state variable that specifies the age of a source~$1$ update currently in service, if any. When a source $1$ update is delivered at time $t$, the age $x_0(t)$ will be reset to $x_1(t)$. In the following, when we refer to age, we specifically mean the age of source $1$ updates at the monitor.

In using a piecewise linear SHS for AoI, the elements of $\bv_q$ will be binary. We will see that the ones in $\bv_q$ correspond to certain relevant components of $\xv(t)$ that grow at unit rate in state $q$ while the zeros mark components of $\xv(t)$ that are irrelevant in state $q$ to the age process and need not be tracked.
For tracking of the age process, the transition reset maps are binary: $\Amat_l\in\set{0,1}^{(n+1)\times (n+1)}$. The set of linear mappings $\set{\Amat_l}$ will depend on the specific queue discipline, and the indexing scheme for updates in the system.
 
\begin{definition}
An age-of-information SHS $(\Qcal,\Bcal,\Acal)$ is an SHS in which the discrete state $q(t)\in\Qcal$ is a continuous-time Markov chain with transitions $l\in\Lcal$ from state $q_l$ to $q'_l$ at rate $\laml$ and the continuous state evolves according to $\dot{\xv}(t)=\bv_q\in\set{0,1}^{n+1}$ in each discrete state $q\in\Qcal$ and is subject to the linear transition reset map $\xv'=\xv\Amat_l$ in transition $l$. 
\end{definition}

Note that column $j$, $0\le j \le n$, of $\Amat_l$ determines how $x'_{j}$ is set when transition $l$ takes place. Typically, we will construct transition mappings $\Amat_l$ that have  no more than a single $1$ in each column. In particular if $[\Amat_l]_{i,j}=1$, then transition $l$ firing causes $x'_j=x_i$. This corresponds to transition $l$ relabeling update $i$ as update $j$. This occurs, for example, in a FCFS queue when the service completion of update $1$ causes update $i$ occupying queue position $i$ to be relabeled as update $i-1$, because its queue position changes to $i-1$. Another important case occurs when a transition $l$ inserts a fresh source $1$ update in queue position $j$ at time $t$. Immediately following the transition, i.e. after that update is inserted, $x'_{j}=0$ because that update is fresh. In this case, 
$[\Amat_l]_j$ is an all-zero column.

On the other hand, if transition $l$ corresponds to the service completion of a source $1$ update indexed $j$, then $\Amat_l$ must encode the resulting age reduction. This would require $[\Amat_l]_{j,0}=1$ and, for $k\ne j$, $[\Amat_l]_{k,0}=0$, so that the mapping $\xv'=\xv\Amat_l$ yields $x'_0=x_{j}$. That is, the age is reset to the age of this most recently delivered update and the corresponding reduction in age is $x_0-x'_0=x_0-x_{j}$.

We note that the SHS method specifies the continuous state $\xv(t)$ for all discrete states $q\in\Qcal$.  However, not all components of $\xv(t)$ are relevant in all states. 
Since the dimensionality of $\xv(t)$ is fixed to be $n+1$, we choose $n$ to be the maximum number of updates in the system over all states $q$. However, when a state $q$ has $k_q<n$ updates in the system, then $x_{k_q+1}(t),\ldots,x_n(t)$ are irrelevant variables in state $q$ as there are no corresponding updates in the system that could complete service. We also note that not all updates in the system are relevant to the future trajectory of the age process.  Since we are tracking the age of source $1$ updates, $x_{j}(t)$ is irrelevant when there is an update from source $s >1$ in queue position $j$.

We define $\Ical_q$ as the index set of irrelevant variables in discrete state $q$. That is, $j\in\Ical_q$ if $x_j(t)$ is irrelevant in state $q$. We will see that the irrelevant variables in discrete state $q$ have no impact on the subsequent state in a transition out of state $q$. Hence, when the system enters state $q$, we can arbitrarily set an irrelevant variable to any value.  For algebraic clarity, we adopt the convention that each irrelevant $x_j(t)$ is zero in state $q$. 
For each state $q$, we set
\begin{align}
[\Amat_l]_{j} &=\zerov[n]^\top,\qquad l\in\Lcal'_q, j\in\Ical_q,
\eqnlabel{irrelevant-A}\\
\shortintertext{and}
[\bv_q]_{j}&=0,\qquad j\in\Ical_q.\eqnlabel{irrelevant-b}
\end{align}
When transition $l$ occurs and state $q$  is entered with $j\in\Ical_q$, \eqnref{irrelevant-A} implies $x_j(t)$ is reset  to $x'_j=0$ and \eqnref{irrelevant-b} ensures $x_j(t)$ remains zero while in state $q$. On the other hand, if in state $q$ the update $j$ is of source $1$, then $x_{j}(t)$ is relevant in state $q$. As this relevant update ages at unit rate in state $q$, we set $[\bv_q]_{j}=1$ for $j\not\in\Ical_q$. 



A foundational assumption for age analysis is that the Markov chain $q(t)$ is ergodic; otherwise, time-average age analysis makes little sense. Under this assumption, the state probability vector $\piv(t)=\rvec{\pi_0(t) &\cdots&\pi_m(t)}$ always  converges to the unique stationary vector $\bar{\piv}=\rvec{\pibar_0&\cdots&\pibar_m}$ satisfying
\begin{subequations}
\begin{align}
\bar{\pi}_{\qbar}\sum_{l\in\Lcal_{\qbar}}\laml&=\sum_{l\in\Lcal'_{\qbar}}\laml
\bar{\pi}_{\ql},\eqnlabel{AOI-SHS-pi}\\
\sum_{\qbar\in\Qcal}\bar{\pi}_\qbar&=1.
\end{align}
\end{subequations}
 Moreover, we see in Lemma~\ref{lem:pi-vv-derivs} that convergence  to $\bar{\piv}$ is disconnected from the evolution of the age process. This is as expected since the age  process $\xv(t)$ is a measurements process that does not influence the evolution of the queue state.  
 
When $\piv(t)$ has converged to the stationary probability vector $\bar{\piv}$,  we see from Lemma~\ref{lem:pi-vv-derivs} that \eqnref{vv-derivs}  is reduced to a system of first order differential equations 
\begin{align}
\dot{\vv}_{\qbar}(t)&=
\bv_{\qbar}\pibar_{\qbar}+\sum_{l\in\Lcal'_{\qbar}}\laml \vv_{\ql}(t)\Amat_l
-\vv_{\qbar}(t)\sum_{l\in\Lcal_{\qbar}}\laml,\quad \qbar\in\Qcal,
\eqnlabel{vv-derivs-pibar}
\end{align}
in $\vv(t)=\rvec{\vv_0(t)&\cdots&\vv_m(t)}$.
%
While Lemma~\ref{lem:pi-vv-derivs} holds for any set of reset maps $\set{\Amat_l}$, the differential equation \eqnref{vv-derivs-pibar} may or may not be stable. Stability depends on the collection of reset maps\footnote{For example, it would be easy to construct reset maps such that $x_0(t)=t$, i.e. $x_0(t)$ simply tracks the passage of time and $v_{q0}(t)$ grows without bound for all states $q\in \Qcal$.}.
When \eqnref{vv-derivs-pibar} is stable, 
each $\vv_{\qbar}(t)=\E{\xv(t)\dqt{\qbar}}$ converges to a limit $\vvbar_{\qbar}$ as $t\goes\infty$. 
In this case, it then follows from \eqnref{x-decomposition} that
\begin{align}
\E{\xv}&=\limty{t}\E{\xv(t)}
\TWOCOLUMN{\nn &}=\limty{t}\sum_{\qbar\in\Qcal} \E{\xv(t)\delta_{\qbar,q(t)}}
=\sum_{\qbar\in\Qcal} \bar{\vv}_{\qbar}
\end{align}
and that the average age of the process of interest is then
$\age=\E{x_0}=\sum_{\qbar\in\Qcal}\vbar_{\qbar 0}$.
We can calculate these limiting values by setting the derivatives  $\dot{\pi}_{\qbar}(t)$ and $\dot{\vv}_q(t)$ in Lemma~\ref{lem:pi-vv-derivs}  to zero and solve for the limiting values $\bar{\pi}_{\qbar}$ and $\bar{\vv}_{\qbar}$.  A consequence is the following theorem. 
\begin{theorem}\thmlabel{AOI-SHS}
If the discrete-state Markov chain $q(t)$ is ergodic with stationary distribution $\bar{\piv}$ and we can find a non-negative solution $\vvbar=\rvec{\vvbar_0&\cdots\vvbar_m}$ 
such that 
\begin{subequations}
\begin{align}
\bar{\vv}_{\qbar}\sum_{l\in\Lcal_{\qbar}}\laml &=\bv_{\qbar}\bar{\pi}_{\qbar}+ \sum_{l\in\Lcal'_{\qbar}}\laml \bar{\vv}_{\ql}\Amat_l,\qquad \qbar\in\Qcal,\eqnlabel{AOI-SHS-v}
\end{align}
then the differential equation \eqnref{vv-derivs-pibar} is stable and the average age of the AoI SHS is given by
\begin{equation}
\age=\sum_{\qbar\in\Qcal} \vbar_{\qbar 0}.
\eqnlabel{age-vsum}
\end{equation}
\end{subequations}
\end{theorem}
Proof of \Thmref{AOI-SHS} is deferred to Section~\ref{sec:SHS-matrix} as some elements of the proof will be more clear after we employ the LCFS-S and LCFS-W systems as examples.  In particular, Section~\ref{sec:LCFS-SHS}  uses   \Thmref{AOI-SHS}  to prove \Thmref{LCFS}.  We will see that  the construction of a simple table that enumerates the transitions $a_l=(\ql,\ql',\laml,\Amat_l)$ will be sufficient to immediately write down and solve the equations of \Thmref{AOI-SHS}. 

We note that \Thmref{AOI-SHS} is in a form convenient for deriving closed form AoI expressions for simple queues. However, this form is not concise in that multiple instances of $\vv_{\ql}$ on the right side of \eqnref{AOI-SHS-v} may refer to the same $\vv_q$.  In Section~\ref{sec:SHS-matrix}, we rewrite these equations in matrix form that is convenient for numerical evaluation and also facilitates a proof of \Thmref{AOI-SHS}. For the non-negative linear reset maps $\Amat_l$ that we employ for age analysis, we will show that stability of the differential equation \eqnref{vv-derivs-pibar} is equivalent  to an eigenvalue constraint on a non-negative matrix. 

\section{LCFS Age: SHS Analysis}\label{sec:LCFS-SHS}
\subsection{LCFS With Preemption In Service: SHS Analysis}
\label{sec:lcfsWithpre-SHS1}
Without loss of generality, we assume a two-source LCFS-S system and we solve for the average age $\age_1$ of source $1$. In terms of the $N$ source system, source $2$ represents the composition of all other sources. We can represent the LCFS-S system discrete state $q(t)=q\in\Qcal=\set{0,1,2}$ such that $q=0$ indicates that the server is idle and $q\in\set{1,2}$ denotes the source of the update packet in service.  The continuous state is $\xv(t)=\cvec{x_0(t)\ x_1(t)}$ where $x_0(t)$ is the current age $\age_1(t)$ of the source 1 process, and $x_1(t)$ encodes what $\age_1(t)$ will become if the packet-in-service is delivered.  We note that $x_1(t)$ is irrelevant in state $0$. In state $1$, $x_1(t)$ is the age of the source $1$ update in service. In state $2$, a source $2$ update is in service. Because a service completion by this update has no effect on the source $1$ age, $x_1(t)$ is also irrelevant in state $2$.
\begin{figure}[t]
\centering
\begin{tikzpicture}[->, >=stealth', auto, semithick, node distance=2.75cm]
\tikzstyle{every state}=[fill=white,draw=black,thick,text=black,scale=1]
\node[state]    (0)                     {$0$};
\node[state]    (1)[below left of=0]   {$1$};
\node[state]    (2)[below right of=0]   {$2$};
\path
(0) 	edge[bend right=20,above]     node{$1$}     	(1)
	edge[bend left=20,above]     node{$2$}     	(2)
(1) edge[bend right=20,right,above] node{$3$} (0)
     edge[loop left,left] 	node{$4$} 	(1)
     edge[bend left=20,above] node{$5$} (2)
 (2) edge[bend left=20,above] node{$6$} (0)
 edge[bend left=20,above] node{$7$} (1);
\end{tikzpicture}
\caption{The SHS Markov chain for updates of source $1$ in the two-source LCFS-S system. In state $0$ the system is idle while in state $i\in\set{1,2}$ a source~$i$ update is in service. The transition rates and transition/reset maps for links $l=1,\ldots,7$ are shown in Table~\ref{tab:MC-LCFS-S}.}
\label{fig:MC-LCFS-S}
\end{figure}
\begin{table}[t]
\begin{displaymath}
\begin{array}{rrrccc}
l & q_l\to q'_l & \laml & \xv\Amat_l &\Amat_l&\vv_{q_l}\Amat_l\\\hline
1 & 0\to 1 &\lambda_1 	& \rvec{x_0& 0}&\smvec{1 & 0\\ 0 & 0}
&\rvec{v_{00}&0}\\[1em]
2 & 0\to 2 & \lambda_2	& \rvec{x_0&0}&\smvec{1 & 0\\ 0 & 0}
&\rvec{v_{00}& 0}\\[1em]
3 & 1\to0& \mu		& \rvec{x_1&0}&\smvec{0 & 0\\ 1 & 0}
&\rvec{v_{11}&0}\\[1em]
4 & 1\to1& \lambda_1 	& \rvec{x_0&0}&\smvec{1 & 0\\ 0 & 0}
& \rvec{v_{10}&0}\\[1em]
5 & 1\to2& \lambda_2& \rvec{x_0&0}&\smvec{1 & 0\\ 0 & 0}
&\rvec{v_{10}&0}\\[1em]
6 & 2\to0&  \mu 		& \rvec{x_0&0}&\smvec{1 & 0\\ 0 & 0}
&\rvec{v_{20}&0}\\[1em]
7 & 2\to1& \lambda_1	&\rvec{x_0&0} &\smvec{1 & 0\\ 0 & 0}
&\rvec{v_{20}&0}
\end{array}
\end{displaymath}
\caption{Table of transitions for the Markov chain in Figure~\ref{fig:MC-LCFS-S}.}\label{tab:MC-LCFS-S}
\end{table}

A Markov chain for the discrete state $q(t)$ is shown in Figure~\ref{fig:MC-LCFS-S}.  The corresponding SHS  transitions $a_l$  are shown in Table~\ref{tab:MC-LCFS-S}.  In the figure, a link $l$ from node $q_l$ to $q'_l$ indicates that  transitions in state $q_l$ to state $q'_l$ occur at exponential rate $\lambda^{(l)}$ as given in the table. 
In constructing the table, we first identify the $\xv\to\xv'$ mapping for each transition $l$, from which we infer the matrix $\Amat_l$ such that $\xv'=\xv\Amat_l$. Given $\Amat_l$, it is convenient to also include $\vv_{\ql}\Amat_l$ in the table  to facilitate the use of \Thmref{AOI-SHS}. We now explain each transition $l$:
\begin{description}
\item[$l=1$] A source $1$ update arrives at an empty queue. With this arrival, $x'_0=x_0$ is unchanged because the arrival does not yield an age reduction until it departs.  However $x'_1=0$ because the arriving source $1$ update is fresh and its age is zero at that instant. 
\item[$l=2$] A source $2$ update arrives at an empty queue. The age $x'_0=x_0$ is unchanged because the arrival does not change the age. However $x'_1=0$ because $x_1$ is irrelevant in state~$2$. 
\item[$l=3$] A source $1$ update completes service and is delivered to the monitor. In this transition, $x_0'=x_1$, corresponding to the age being reset to the age of the source $1$ that just completed service. Also note that $x'_1=0$ since $x_1$ becomes irrelevant when the system enters state $0$. 
\item[$l=4$] The source $1$ update in service is preempted by a fresh source $1$ update. The age $x_0$ remains unchanged while $x_1$ is reset to zero because the new update is fresh.
\item[$l=5$] The source $1$ update in service is preempted by a source $2$ update. The age $x'_0=x_0$ is unchanged and $x'_1=0$ since $x_1$ becomes irrelevant in state $2$.
\item[$l=6$] A source $2$ update completes service. The source $1$ age $x_0$ is unchanged.  In the transition to state $0$, $x_1$ remains irrelevant and is set to zero.
\item[$l=7$] The source $2$ update in service is preempted by a fresh source $1$ update. The age $x_0$ is unchanged while $x'_1=0$ because the new update is fresh.
\end{description}


We note that this SHS includes a self-transition in which the discrete state is unchanged but a reset occurs in the continuous state. Specifically, in state $1$, the self-transition of link $4$ marks the arrival of a source $1$ update packet that preempts the source $1$ packet in service. This leaves the discrete state $q(t)$ and the current age $x_0(t)$ unchanged, but the more recent timestamp of the new source $1$ update resets $x_1(t)$. 

The evolution of $\xv(t)$ depends on the discrete state $q(t)$. Specifically, when $q(t)=q$,
\begin{equation}\eqnlabel{xderiv-S}
\dot{\xv}(t)=\bv_{q}=\begin{cases}
\rowvec{1&0} & q=0,2,\\
\rowvec{1&1} & q=1.
\end{cases}
\end{equation}
The interpretation  of \eqnref{xderiv-S} is that  the age $\age_1(t)=x_0(t)$ increases at unit rate with time $t$ in all discrete states but $x_1(t)$ increases at unit rate only in state $q=1$ in which there is a relevant update in service.

\mbox{}To employ \Thmref{AOI-SHS}, we first use  \eqnref{AOI-SHS-pi} to show that the stationary probability vector $\bar{\piv}$ satisfies  $\bar{\piv}\Dmat=\bar{\piv}\Qmat$
with
\begin{align}
\Dmat&=\diag{\lambda,\mu+\lambda,\mu+\lambda_1}, &
\Qmat&=
\begin{bmatrix}
0 & \lambda_1 & \lambda_2\\
\mu & \lambda_1 & \lambda_2\\
\mu & \lambda_1 & 0\end{bmatrix}.
\end{align}
Applying 
$\sum_{i=0}^2\pibar_i=1$, the  stationary probabilities are
\begin{align}\eqnlabel{S-statprobs}
\begin{bmatrix}
\pibar_0 & \pibar_1 &\pibar_2\end{bmatrix}&=(1+\rho)^{-1}\begin{bmatrix}1&\rho_1&\rho_2\end{bmatrix}.
\end{align}

Since $\bv_0=\rowvec{1 & 0}$, evaluation of \eqnref{AOI-SHS-v} in \Thmref{AOI-SHS} at $\qbar=0$ yields
\begin{align}\eqnlabel{LCFS-S-qbar0}
\lambda\rvec{\vbar_{00}&\vbar_{01}}
=\rvec{\pibar_0&0} +\mu\rvec{\vbar_{11}&0}
+\mu\rvec{\vbar_{20}&0}.
\end{align}
We see from \eqnref{LCFS-S-qbar0}  that $\vbar_{01}=0$. This is a consequence of $x_1$ being irrelevant in state $q=0$.  In particular, $x_1(t)\dqt{0}=0$ for all $t$ because $x_1(t)$ is held at $0$ when $q(t)=0$ (by our convention for irrelevant variables) and  $\dqt{0}=0$ when $q(t)\neq 0$.
Thus $v_{01}(t)=\E{x_1(t)\dqt{0}}=0$ and we refer to $v_{01}$ as irrelevant.  In general, when $x_j$ is irrelevant in state $q$, $v_{qj}(t)=0$ for all $t$ and we refer to $v_{qj}(t)$ as irrelevant.

Evaluating \eqnref{AOI-SHS-v} at $\qbar=1$ and $\qbar=2$ produces
\begin{subequations}
\eqnlabel{LCFS-S-qbar12}
\begin{align}
(\mu+\lambda)\rvec{\vbar_{10}&\vbar_{11}}&=\rvec{\pibar_1& \pibar_1}
+\lambda_1\rvec{\vbar_{00} & 0}+\lambda_1\rvec{\vbar_{10} & 0}
+\lambda_1\rvec{\vbar_{20} & 0} 
\eqnlabel{LCFS-S-qbar1},\\
(\mu+\lambda_1)\rvec{\vbar_{20}&\vbar_{21}}&=\rvec{\pibar_2&0}+\lambda_2\rvec{\vbar_{00}&0} 
+\lambda_2\rvec{\vbar_{10}&0}.\eqnlabel{LCFS-S-qbar2}
\end{align}
\end{subequations}
In terms of the vectors 
\begin{align}
\bar{\piv}&=\rvec{\pibar_0&\pibar_1&\pibar_2},\\
\vvbar&=\rowvec{\vvbar_0 &\vvbar_1 &\vvbar_2}=\rvec{\vbar_{00}&\vbar_{01}&\vbar_{10}&\vbar_{11}&\vbar_{20}&\vbar_{21}},
\end{align}
we have
\begin{align}\eqnlabel{DBR-LCFS-S}
\vvbar\Dmat&=\bar{\piv}\Bmat+\vvbar\Rmat
\end{align}
where
\begin{align}
\Dmat&=\diag{\lambda,\lambda,\mu+\lambda,\mu+\lambda,\mu+\lambda_1,\mu+\lambda_1},\\
\Bmat &=\begin{bmatrix}
1 & 0 & 0 & 0 & 0 & 0\\
0 & 0 & 1 & 1 & 0 & 0\\
0 & 0 & 0 & 0 & 1 & 0
\end{bmatrix},
& \Rmat&=\begin{bmatrix}
0    & 0    & \lambda_1 & 0 & \lambda_2 & 0\\
0    & 0    &  0              & 0 & 0              & 0\\
0    & 0    & \lambda_1 & 0 & \lambda_2 & 0\\
\mu& 0    & 0               & 0 & 0              & 0\\
\mu& 0    & \lambda_1 & 0 &  0             & 0\\
0    & 0    &  0              & 0 & 0              & 0
\end{bmatrix}.\eqnlabel{BR-LCFS-S}
\end{align}
We observe that the columns and  rows of $\Rmat$ corresponding to the irrelevant variables $\vbar_{01}$ and $\vbar_{21}$ are zero. Gathering the relevant  variables, we obtain
\begin{align}\eqnlabel{S-Veqns}
\frac{1}{\mu}\rowvec{\pibar_0 & \pibar_1 &\pibar_1&\pibar_2}
&=\rowvec{\vbar_{00} & \vbar_{10} & \vbar_{11} & \vbar_{20}}
\begin{bmatrix}
\rho & -\rho_1   & 0       & -\rho_2\\
0      & 1+\rho_2  &  0      & -\rho_2\\
-1     & 0           & 1+\rho & 0\\
-1     & -\rho_1  & 0       &1+\rho_1
\end{bmatrix}.
\end{align}
It follows from \eqnref{S-statprobs} and \eqnref{S-Veqns} that 
\begin{subequations}\eqnlabel{S-V012}
\begin{align}
\vbar_{00} &= \frac{1}{\mu(1+\rho)}
\bracket{\frac{1+\rho_2}{\rho_1}+\frac{1}{1+\rho}},\\
\vbar_{10} &=\frac{1}{\mu(1+\rho)}
\bracket{1+\rho+\frac{\rho_1}{1+\rho}},\\
\vbar_{20} &=\frac{1}{\mu(1+\rho)}\bracket{\frac{\rho_2(1+\rho)}{\rho_1}+\frac{\rho_2}{1+\rho}}.
\end{align}
\end{subequations}
From \eqnref{LCFS-S-qbar12}, it can be seen that $\vbar_{11}$ is also non-negative. Thus,  \Thmref{AOI-SHS} implies that the average age for source~$1$ is $\age=\sum_{q=0}^2\vbar_{q0}$. Applying  \eqnref{S-V012} yields \Thmref{LCFS}(a) for source $i=1$.
%

\subsection{LCFS-S: A simpler SHS analysis}\label{sec:lcfsWithpre-SHS2}
We note that the preceding analysis of the LCFS-S system used the discrete state to track the source of the update in service. However, since all updates are served at rate $\mu$, specifying the source  of an update in service is not essential for a Markov model for the server occupancy. We will now show that we can track the source $1$ age without specifying the source  of the update in service. 
By employing transition reset maps that depend on the source of the arriving new update, we can forgo using the discrete state to track the source of an update in service. We now demonstrate this technique with a simpler SHS derivation of the LCFS-S age. In this analysis, the discrete state tracks only whether the server is busy. 
 
\begin{figure}[t]
\centering
\begin{tikzpicture}[->, >=stealth', auto, semithick, node distance=2.75cm]
\tikzstyle{every state}=[fill=white,draw=black,thick,text=black,scale=1]
\node[state]    (0)                     {$0$};
\node[state]    (1)[right of=0]   {$1$};
\path
(0) 	edge[bend left=20,above]     node{$1$}     	(1)
	edge[bend left=60,above]     node{$2$}     	(1)
(1) edge[bend left=20,right,below] node{$3$} (0)
     edge[loop above] 	node{$4$} 	(1)
     edge[loop right] 	node{$5$} 	(1);
\end{tikzpicture}
\caption{The simplified SHS Markov chain  for updates of source $1$  in the two-source LCFS-S system. In state $0$ the system is idle while in state $1$ an update of either source $1$ or $2$ is in service. The transition rates and transition/reset maps for links $l=1,\ldots,5$ are shown in Table~\ref{tab:MC-LCFS-S2}.}
\label{fig:MC-LCFS-S2}
\end{figure}
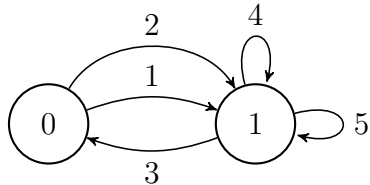
\begin{table}[t]
\begin{displaymath}
\begin{array}{rrrccc}
l & q_l\to q'_l & \laml & \xv\Amat_l &\Amat_l&\vv_{q_l}\Amat_l\\[1mm]\hline\\[-4mm]
1 & 0\to 1 &\lambda_1 	& \rvec{x_0& 0}&\smvec{1 & 0\\ 0 & 0}
&\rvec{v_{00}&0}\\[1em]
2 & 0\to 1 & \lambda_2	& \rvec{x_0&x_0}&\smvec{1 & 1\\ 0 & 0}
&\rvec{v_{00}&v_{00}}\\[1em]
3 & 1\to0& \mu		& \rvec{x_1&0}&\smvec{0 & 0\\ 1 & 0}
&\rvec{v_{11}&0}\\[1em]
4 & 1\to1& \lambda_1 	& \rvec{x_0&0}&\smvec{1 & 0\\ 0 & 0}
& \rvec{v_{10}&0}\\[1em]
5 & 1\to1& \lambda_2& \rvec{x_0&x_0}&\smvec{1 & 1\\ 0 & 0}
&\rvec{v_{10}&v_{10}}
\end{array}
\end{displaymath}
\caption{Table of transitions for the Markov chain in Figure~\ref{fig:MC-LCFS-S2}.}\label{tab:MC-LCFS-S2}
\end{table}

 Just as in the original LCFS-S SHS, the continuous state is $\xv(t)=\rowvec{x_0(t) & x_1(t)}$ where $x_0$ is the current source $1$ age. However, $x_1(t)$ is now what the age would become if the update in service is delivered. In state $1$, both $x_0(t)$ and $x_1(t)$ increase at unit rate; i.e. $\bv_1=\rowvec{1&1}$. On the other hand, in state $0$, $x_1$ is meaningless and $\bv_0=\rowvec{1 & 0}$. The transitions are:
 \begin{description}
 \item[$l=1$] A fresh source $1$ update goes into service; $x_1'=0$ because the update is fresh. 
 \item[$l=2$] A fresh source $2$ update goes into service and $x'_1=x_0$. If the source $2$  update does complete service, it doesn't reduce the age of the process of interest.
 \item[$l=3$] The update in service is delivered. The age $x_0$ is reset to $x'_0=x_1$. If this delivered update is from source $1$, then  $x'_0<x_0$. However, if this update is from source $2$, then $x_0'=x_0$ and no age reduction occurs. Note that this age reduction was encoded in the prior transition that put this update in service.
 \item[$l=4$] The update in service is replaced by a fresh source $1$ update. This reset map is essentially the same as for transition $l=1$.
 \item[$l=5$] The update in service is replaced by a fresh source $2$ update. This reset map is essentially the same as for transition $l=2$.
\end{description}
The Markov chain for the discrete state has stationary probabilities 
\begin{equation}
{\pivbar=\rowvec{\pibar_0&\pibar_1}}=\vec{\frac{1}{1+\rho} & \frac{\rho}{1+\rho}}.
\end{equation}
In this system, $\vv_0=\rowvec{v_{00} & v_{01}}$ and $\vv_1=\rowvec{v_{10} & v_{11}}$. Evaluating \eqnref{AOI-SHS-v} at $\qbar=0,1$  produces
\begin{subequations}
\begin{align}
\lambda\rvec{\vbar_{00}&\vbar_{01}}
&=\rvec{\pibar_0&0} +\mu\rvec{\vbar_{11}&0},
\eqnlabel{LCFS-S2-qbar0}\\
(\mu+\lambda)\rvec{\vbar_{10}&\vbar_{11}}&=\rvec{\pibar_1& \pibar_1}
+\lambda_1\rvec{\vbar_{00} & 0}+\lambda_2\rvec{\vbar_{00}&\vbar_{00}}+\lambda_1\rvec{\vbar_{10} & 0}
+\lambda_2\rvec{\vbar_{10} & \vbar_{10}} 
\eqnlabel{LCFS-S2-qbar1}.
\end{align}
\end{subequations}

As expected, we see from \eqnref{LCFS-S2-qbar0}  that $\vbar_{01}=0$ because $x_1$ is irrelevant in state $0$.  Normalizing by the service rate $\mu$, we obtain
\begin{subequations}
\eqnlabel{LCFS-S2-V01}
\begin{align}
\rho\vbar_{00}
&=\pibar_0/{\mu}+\vbar_{11},\\
(1+\rho)\vbar_{10}&={\pibar_1}/{\mu}
+\rho\vbar_{00} +\rho\vbar_{10},\\
(1+\rho)\vbar_{11}&={\pibar_1}/{\mu}
+\rho_2\vbar_{00}+\rho_2\vbar_{10}. 
\end{align}
\end{subequations}
Solving \eqnref{LCFS-S2-V01}, it can be shown that $\vbar_{00}$, $\vbar_{10}$ and $\vbar_{11}$  are all non-negative. Moreover, calculation of $\age=\vbar_{00}+\vbar_{10}$ yet again yields \Thmref{LCFS}(a) for source $i=1$.

\subsection{LCFS-S: An even simpler SHS analysis with fake updates}\label{sec:lcfsWithpre-SHS3}
We note that the preceding analysis of the LCFS-S system used the discrete state to track whether an update is in service.  However, it turns out that this is not essential and we now analyze AoI in the multi-source LCFS-S queue using a one-state SHS in which there is always an update in service.  The key idea is that when an update completes service and the server would become idle, we create a ``fake'' update to keep the server busy. This fake update is a duplicate of the previous update and has its same timestamp. The reason this trick works is two-fold. First, if the fake update were to complete service, the age at the monitor is unchanged because the update timestamp is the same as that of the previously delivered update. Second, when a new (true) update is submitted, it immediately preempts any fake update that may have been keeping the server busy.

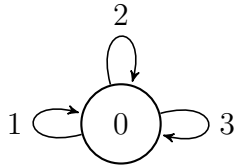
\begin{figure}[t]
\centering
\begin{tikzpicture}[->, >=stealth', auto, semithick, node distance=2.75cm]
\tikzstyle{every state}=[fill=white,draw=black,thick,text=black,scale=1]
\node[state]    (0)                     {$0$};
\path
(0) edge[loop left] 	node{$1$} 	(0)
     edge[loop above] 	node{$2$} 	(0)
     edge[loop right] node{$3$} (0);
\end{tikzpicture}
\caption{The simplified SHS Markov chain  for updates of source $1$  in the two-source LCFS-S system. The system is always busy serving either a real or fake update.  The transition rates and transition/reset maps for links $l=1,2,3$ are shown in Table~\ref{tab:MC-LCFS-S3}.}
\label{fig:MC-LCFS-S3}
\end{figure}
\begin{table}[t]
\begin{displaymath}
\begin{array}{rrrccc}
l & q_l\to q'_l & \laml & \xv\Amat_l &\Amat_l&\vv_{q_l}\Amat_l\\[1mm]\hline\\[-4mm]
1 & 0\to 0 &\lambda_1 	& \rvec{x_0& 0}&\smvec{1 & 0\\ 0 & 0}
&\rvec{v_{00}&0}\\[1em]
2 & 0\to 0 & \lambda_2	& \rvec{x_0&x_0}&\smvec{1 & 1\\ 0 & 0}
&\rvec{v_{00}&v_{00}}\\[1em]
3 & 0\to0& \mu		& \rvec{x_1&x_1}&\smvec{0 & 0\\ 1 & 1}
&\rvec{v_{01}&v_{01}}\\[1em]
\end{array}
\end{displaymath}
\caption{Table of transitions for the Markov chain in Figure~\ref{fig:MC-LCFS-S2}.}\label{tab:MC-LCFS-S3}
\end{table}

The one-state SHS is shown in Figure~\ref{fig:MC-LCFS-S3} and the corresponding table of transitions is given in Table~\ref{tab:MC-LCFS-S3}.  
Just as in the previous LCFS-S SHS, the continuous state is $\xv(t)=\rowvec{x_0(t) & x_1(t)}$ where $x_0$ is the current source $1$ age and $x_1(t)$ is what the age would become if the update in service is delivered. Both $x_0(t)$ and $x_1(t)$ increase at unit rate and $\bv_0=\rowvec{1&1}$.  The transitions are:
 \begin{description}
 \item[$l=1$] A fresh source $1$ update goes into service; $x_1'=0$ because the update is fresh. 
 \item[$l=2$] A fresh source $2$ update goes into service and $x'_1=x_1$. If the source $2$  update does complete service, it doesn't reduce the age of the process of interest.
 \item[$l=3$] The update in service is delivered. The age $x_0$ is reset to $x'_0=x_1$ but $x_1$ is unchanged: $x_1'=x_1$. This corresponds to creating a fake update with the same timestamp as the update that was just delivered.
 \end{description}

The Markov chain for the discrete state has the trivial stationary probability $\pi_0=1$.
In this system, $\vv_0=\rowvec{v_{00} & v_{01}}$. Evaluating \eqnref{AOI-SHS-v} at $\qbar=0$  produces
\begin{align}
(\mu+\lambda)\rvec{\vbar_{00}&\vbar_{01}}
&=\rvec{1 & 1} +\lambda_1\rvec{v_{00}&0}+\lambda_2\rvec{v_{00} & v_{00}} +\mu\rowvec{v_{01} & v_{01}}.
\eqnlabel{LCFS-S3-qbar0}
\end{align}
Solving these two equations for $v_{00}$ and $v_{01}$, the average age $\age=v_{00}$ yet again yields \Thmref{LCFS}(a) for source $i=1$.

\subsection{LCFS With Preemption Only In Waiting: SHS Analysis}
\label{sec:lcfsWOpreMM1}

Following the simplified SHS method introduced in Section~\ref{sec:lcfsWithpre-SHS2}, we now model the LCFS-W system as a  stochastic hybrid system. 
Once again  we assume a two-source system and we solve for the average age $\age_1$ of source $1$. In terms of the $N$ source system, source $2$ represents the composition of all other sources. 

\begin{figure}[t]
\centering
\begin{tikzpicture}[->, >=stealth', auto, semithick, node distance=2.75cm]
\tikzstyle{every state}=[fill=white,draw=black,thick,text=black,scale=1]
\node[state]    (0)                     {$0$};
\node[state]    (1)[right of=0]   {$1$};
\node[state]    (2)[right of=1]   {$2$};
\path
(0) 	edge[bend left=20,above]     node{$1$}     	(1)
	edge[bend left=60,above]     node{$2$}     	(1)
(1) edge[bend left=20,right,below] node{$3$} (0)
     edge[bend left=20,above]     node{$4$}     	(2)
	edge[bend left=60,above]     node{$5$}     	(2)
(2) edge[bend left=20,right,below] node{$6$} (1)
    edge[loop above] 	node{$7$} 	(2)
     edge[loop right] 	node{$8$} 	(2);
\end{tikzpicture}
\caption{The simplified SHS Markov chain  for updates of source $1$  in the two-source LCFS-W system. The state $i$ indicates the number of updates in the system. The transition rates and transition/reset maps for links $l=1,\ldots,8$ are shown in Table~\ref{tab:MC-LCFS-W2}.}
\label{fig:MC-LCFS-W2}
\end{figure}
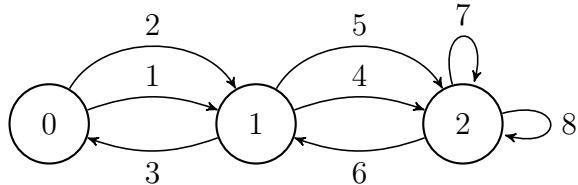

The LCFS-W system  with discrete states $q\in\Qcal=\set{0,1,2}$ is shown in Figure~\ref{fig:MC-LCFS-W2} with the corresponding transition rates $\laml$ and reset maps $\Amat_l$ given in Table~\ref{tab:MC-LCFS-W2}. Much like the two-state analysis of the SHS for the LCFS-S system, the discrete state tracks the number of updates in the system but not the source of each update. Whether a delivered update reduces the age of source $1$ is embedded in the continuous state.\footnote{The fake updates method fails for the LCFS-W system because the discrete state must track whether the update in service is real or fake because an update in service would be preempted if it were fake but not if it were real.}

The continuous state is $\xv(t)=\cvec{x_0(t)\ x_1(t)\ x_2(t)}$ where $x_0(t)$ is the current age $\age_1(t)$ of the source $1$ process, $x_1(t)$ is what the age would be if the update in service were delivered at time $t$, and $x_2(t)$ is what the age would be if the update-in-waiting were delivered at time $t$.  In state $q=0$, $x_1$ and $x_2$ are irrelevant. In state $q=1$, $x_2$ is irrelevant. Following our prior convention, relevant components in each state grow at unit rate while irrelevant components are fixed at zero. Consequently,  in discrete state $q(t)=q$, the continuous state  evolves according to
\begin{equation}\eqnlabel{xderiv-W}
\dot{\xv}(t)=\bv_{q}=\begin{cases}
\rowvec{1&0&0}, & q=0,\\
\rowvec{1&1&0}, & q=1,\\
\rowvec{1&1&1}, & q=2.
\end{cases}
\end{equation}

For $x_0(t)$ and $x_1(t)$, the transition maps are similar to those in the two-state SHS for the LCFS-S system. The additional complications involve how $x_2(t)$ modifies $x_1(t)$ when an update completes service (and a waiting update goes into service) and how $x_2(t)$ is modified when a waiting update is preempted.
In particular, the reset maps $\Amat_5$ and $\Amat_8$ for transitions $l=5$ and $l=8$ are less straightforward than the others. Under transition $l=5$, the update in service is joined by a new source $2$ update that waits in the queue. Assuming it is not preempted in waiting, this new source $2$ update enters service and is eventually delivered only after the update in service is delivered. This delivery of the update in service reduces the age to $x_1$ and puts the new source $2$ update into service. However, as the SHS is tracking the source $1$ age, the eventual delivery of the new source $2$ update 
will not reduce the age. Thus transition $l=5$ sets $x'_2=x_1$ so that the age upon delivery of the new source $2$ update will leave the age unchanged from the age that will be established by the prior service completion. The same effect occurs in transition $l=8$ in which an arriving source $2$ update preempts the update-in-waiting.

\begin{table}[t]
\begin{displaymath}
\begin{array}{ccccc}
l & q_l\to q'_l & \laml &  \xv\Amat_l & \vv_{\ql}\Amat_l\\\hline
1 & 0\to 1 &\lambda_1 	& \rvec{x_0&0&0}&\rvec{v_{00}&0&0}\\
2 & 0\to 1 & \lambda_2	& \rvec{x_0&x_0&0}&\rvec{v_{00}&v_{00}&0}\\
3 & 1\to0& \mu		& \rvec{x_1&0&0}&\rvec{v_{11}&0&0}\\
4 & 1\to2& \lambda_1 	& \rvec{x_0&x_1&0}
&\rvec{v_{10}&v_{11}&0}\\
5 & 1\to2& \lambda_2	& \rvec{x_0&x_1&x_1}
&\rvec{v_{10}&v_{11}&v_{11}}\\
6 & 2\to1&  \mu		& \rvec{x_1&x_2&0}
&\rvec{v_{21}&v_{22}& 0}\\
7 & 2\to2& \lambda_1	& \rvec{x_0&x_1&0}
&\rvec{v_{20}&v_{21}&0}\\
8 & 2\to2& \lambda_2   	& \rvec{x_0&x_1&x_1}
&\rvec{v_{20}&v_{21}&v_{21}}
\end{array}
\end{displaymath}
\caption{Table of transitions for the Markov chain in Figure~\ref{fig:MC-LCFS-W2}.}\label{tab:MC-LCFS-W2}
\end{table} 

From the Markov chain in Figure~\ref{fig:MC-LCFS-W2}, it is easy to see that the discrete state has stationary distribution
\begin{equation}\eqnlabel{LCFS-W-pi}
\rvec{\pibar_0&\pibar_1&\pibar_2}
=C_{\pi}\rvec{1 & \rho & \rho^2}
\end{equation}
where $C_{\pi}=(1+\rho+\rho^2)^{-1}$ is the normalizing constant. 
From \eqnref{AOI-SHS-v} with $\qbar\in\Qcal$, we obtain
\begin{subequations}\eqnlabel{LCFS-W-v}
\begin{align}
\eqnlabel{LCFS-W-v0}
\lambda\vvbar_0
&=\rvec{\pibar_0&0&0}+\mu\rvec{\vbar_{11}&0&0},\\
\eqnlabel{LCFS-W-v1}
(\lambda+\mu)\vvbar_1
&=\rvec{\pibar_1&\pibar_1&0}+\lambda_1\rvec{\vbar_{00}&0&0}
+\lambda_2\rvec{\vbar_{00}&\vbar_{00}&0}+\mu\rvec{\vbar_{21}&\vbar_{22}&0},\\
(\lambda+\mu)\vvbar_2
&=\rvec{\pibar_2&\pibar_2&\pibar_2}+\lambda_1\rvec{\vbar_{10}&\vbar_{11}&0}+\lambda_2\rvec{\vbar_{10}&\vbar_{11}&\vbar_{11}}\nn
&\quad\qquad\qquad\qquad+\lambda_1\rvec{\vbar_{20}&\vbar_{21}&0}
+\lambda_2\rvec{\vbar_{20}&\vbar_{21}&\vbar_{21}}.
\end{align}
\end{subequations}
We see from \eqnref{LCFS-W-v0} that $\vbar_{01}$ and  $\vbar_{02}$ are zero because $x_1(t)$ and $x_2(t)$ are irrelevant in state $0$. Similarly, \eqnref{LCFS-W-v1} implies $\vbar_{12}=0$ because $x_2(t)$ is irrelevant in state $1$.
Gathering the relevant variables and normalizing by the service rate $\mu$, we obtain 
\begin{subequations}
\eqnlabel{LCFS-W-veqns}
\begin{align}
\rho \vbar_{00} &=\pibar_0/\mu +\vbar_{11},\\
(1+\rho)\vbar_{10}&=\pibar_1/\mu +\rho\vbar_{00} +\vbar_{21},\\
(1+\rho)\vbar_{11} &=\pibar_1/\mu+\rho_2\vbar_{00}+\vbar_{22},\\
\vbar_{20}&=\pibar_2/\mu +\rho\vbar_{10},\\
\vbar_{21}&=\pibar_2/\mu +\rho \vbar_{11}\eqnlabel{LCFS-W-v22},\\
(1+\rho)\vbar_{22} &=\pibar_2/\mu+ \rho_2\vbar_{11} +\rho_2\vbar_{21}.\eqnlabel{LCFS-W-v23}
\end{align}
\end{subequations}
We employ \eqnref{LCFS-W-v22} and \eqnref{LCFS-W-v23} to write
\begin{equation}
\vbar_{22}=\frac{1+\rho_2}{1+\rho}\pibar_2+\rho_2\vbar_{11}.
\eqnlabel{LCFS-W-v23a}
\end{equation}
We now apply \eqnref{LCFS-W-v22} and \eqnref{LCFS-W-v23a} to the other equations in  \eqnref{LCFS-W-veqns}, yielding
\begin{subequations}
\eqnlabel{LCFS-W-veqns2}
\begin{align}
\rho \vbar_{00} &=\pibar_0 +\vbar_{11},\\
\vbar_{10}&=\frac{1}{\mu(1+\rho)}+\vbar_{11},\\
(1+\rho)\vbar_{11} &=\pibar_1/\mu+\rho_2\vbar_{00}+\frac{1+\rho_2}{\mu(1+\rho)}\pibar_2+\rho_2\vbar_{11},\\
\vbar_{20}&=\pibar_2/\mu +\rho\vbar_{10}.
\end{align}
\end{subequations}
From \eqnref{LCFS-W-veqns2}, some algebra will show
\begin{equation}
\vbar_{11}=\frac{\rho}{\mu(1+\rho)}\frac{1}{\rho_1} -\frac{C_{\pi}(1+\rho+\rho^3)}{\mu(1+\rho)^2}.
\eqnlabel{LCFS-W-v12}
\end{equation}
To verify that $\vbar_{11}$ is non-negative, we note that $\rho_1\le \rho$ and that for fixed $\rho$, $\vbar_{11}$ is minimized over all $\rho_1$ at $\rho_1=\rho$. Some algebra will verify that $\vbar_{11}\ge0$ when $\rho_1=\rho$. It then follows from \eqnref{LCFS-W-veqns} and \eqnref{LCFS-W-veqns2} that all components of $\vvbar$ are non-negative. 
Moreover, it also follows from \eqnref{age-vsum} and \eqnref{LCFS-W-veqns2} that the average age is
\begin{align}
\age&=\vbar_{00} +\vbar_{10}+\vbar_{20}\nn
&=\frac{1}{\mu}+\frac{\pi_0+\pi_2}{\rho}+\frac{1+\rho+\rho^2}{\rho}\vbar_{11}.
\eqnlabel{LCFS-W-age}
\end{align}
The claim of \Thmref{LCFS}(b) then follows from substitution of \eqnref{LCFS-W-pi} and \eqnref{LCFS-W-v12} in \eqnref{LCFS-W-age}.

\section{SHS Matrix Reformulation}\label{sec:SHS-matrix}
In this section, we derive a matrix representation of \eqnref{vv-derivs-pibar} as well as \eqnref{AOI-SHS-v} in \Thmref{AOI-SHS}. This reformulation will facilitate a proof of \Thmref{AOI-SHS}.
Starting with the differential equations \eqnref{vv-derivs-pibar}, we define the departure rate from state $\qbar$ as 
\begin{equation}
d_{\qbar}=\sum_{l\in\Lcal_{\qbar}}\laml. 
\end{equation}
We also define 
\begin{align}\eqnlabel{Lcalij}
\Lcal_{ij}&=\set{l\in\Lcal: q_l=i,q'_l=j},\quad  i,j\in\Qcal,
\end{align}
as the set of SHS transitions from state $i$ to state $j$.  With the observation that $\Lcal'_{\qbar}=\cup_{i}\Lcal_{i\qbar}$, we now can rewrite  \eqnref{vv-derivs-pibar} as
\begin{align}
\dot{\vv}_{\qbar}(t)&=
\bv_{\qbar}\pibar_{\qbar}+\sum_i\sum_{l\in\Lcal_{i\qbar}}\laml \vv_{\ql}(t)\Amat_l
-d_{\qbar}\vv_{\qbar}(t),\quad \qbar\in\Qcal.\eqnlabel{vv-derivs-pibar2}
\end{align}
With the substitution $\qbar=j$ and the observation that $\ql=i$ for all $l\in\Lcal_{ij}$,  we obtain
\begin{align}
\dot{\vv}_{j}(t)&=
\bv_{j}\pibar_{j}+\sum_i\vv_i(t)\sum_{l\in\Lcal_{ij}}\laml\Amat_l
-d_{j}\vv_{j}(t),\quad j\in\Qcal.\eqnlabel{vv-derivs-pibar3}
\end{align}
We define the block matrix $\Rmat$ such that block $i,j$ of $\Rmat$ is given by
\begin{align}
\Rmat_{ij} &=  \sum_{l\in\Lcal_{ij}}\laml \Amat_l,\qquad i,j\in\Qcal.
\end{align}
We also define the block diagonal matrices\footnote{Note that $\Bmat$ is an $(m+1)\times (m+1)(n+1)$ matrix with $i$th row
$\rowvec{\zerov[(n+1)i]&\bv_i&\zerov[(n+1)(m-i)]}$, $i=0,1,\ldots,m$.}
\begin{align}
\Bmat&=\diag{\bv_0,\bv_1,\ldots,\bv_m},\\
\Dmat&=\diag{d_0\Imat_n,d_1\Imat_n,\ldots,d_m\Imat_n}.
\end{align}
With the definition of the long row vector $\vv(t)=\rvec{\vv_0(t) & \cdots&\vv_m(t)}$, we can write \eqnref{vv-derivs-pibar3} in vector form as
\begin{align}\eqnlabel{vvde-matrix}
\dot{\vv}(t) =\bar{\piv}\Bmat+\vv(t)(\Rmat-\Dmat).
\end{align}
We note that setting $\dot{\vv}(t)=\zerov$ and solving for $\vv(t)=\vvbar$ yields 
\begin{align}\eqnlabel{DBR}
\vvbar\Dmat&=\bar{\piv}\Bmat+\vvbar\Rmat,
\end{align}
just as we observed  in \eqnref{DBR-LCFS-S} for the LCFS-S system. 

We note that \eqnref{vvde-matrix} and \eqnref{DBR} are vectorized forms of \eqnref{vv-derivs-pibar} and \eqnref{AOI-SHS-v}.
In vector form, the claim of \Thmref{AOI-SHS} is that  a non-negative solution $\vvbar$ for \eqnref{DBR} implies the differential equation \eqnref{vvde-matrix} is stable and thus $\E{x_0(t)}$ converges to the average age. 

As we saw in \eqnref{BR-LCFS-S}  for the LCFS-S example, there may be irrelevant variables that yield zero columns in $\Bmat $ and corresponding zero rows and zero columns in $\Rmat$. These irrelevant variables will have zero derivatives and will be perpetually zero. Stability of the differential equations depends only on the stability of the relevant variables.  Thus we omit the irrelevant variables and form $\hat{\vv}$, a long row vector of the relevant variables in $\vvbar$. Similarly, we form $\hat{\Bmat}$ and $\hat{\Dmat}$ by deleting the rows/columns of $\Bmat$ and $\Dmat$ corresponding to irrelevant variables. It follows that relevant variables satisfy
\begin{align}\eqnlabel{vvde-matrix-relevant}
\frac{d\hat{\vv}(t)}{dt} =\bar{\piv}\hat{\Bmat}+\hat{\vv}(t)(\hat{\Rmat}-\hat{\Dmat}).
\end{align}
If there is a non-negative solution $\vvbar$ for \eqnref{DBR}, then the relevant components have a fixed point $\hat{\vv}(t)=\tilde{\vv}$ that satisfies
 \begin{align}\eqnlabel{DBRtilde}
\tilde{\vv}\hat{\Dmat}&=\bar{\piv}\hat{\Bmat}+\tilde{\vv}\hat{\Rmat}.
\end{align}
 Let $s=\max_i d_i$, then $s\Imat -\hat{\Dmat}$ is a non-negative diagonal matrix.  Adding $\tilde{\vv}(s\Imat-\hat{\Dmat})$ to both sides of \eqnref{DBRtilde} yields
 \begin{align}\eqnlabel{DBR-uniformized}
\tilde{\vv}s
&=\bar{\piv}\hat{\Bmat}+\tilde{\vv}(s\Imat+\hat{\Rmat}-\hat{\Dmat}).
\end{align}
Because the reset maps $\Amat_l$ are binary, the matrices $\Rmat$ and $\hat{\Rmat}$ are non-negative and  thus the matrix $s\Imat+\hat{\Rmat}-\hat{\Dmat}$ is also non-negative. It follows that  $s\Imat+\hat{\Rmat}-\hat{\Dmat}$ has a dominant real eigenvalue $r(s)\ge0$ with an associated non-negative non-zero right eigenvector $\uv^\top$ such that $\abs{\epsilon}\le r(s)$ for any other eigenvalue $\epsilon$  \cite[Exercise 1.12]{seneta}\footnote{This is a weak form of the Perron-Frobenius theorem that does not require irreducibility of the non-negative matrix.}. Right multiplying \eqnref{DBR-uniformized} by $\uv^\top$, we obtain
\begin{align}
s\tilde{\vv}\uv^\top=\bar{\piv}\hat{\Bmat}\uv^\top 
+r(s)\tilde{\vv}\uv^\top,
\end{align}
which simplifies to
\begin{align}
[s-r(s)]\tilde{\vv}\uv^\top=\bar{\piv}\hat{\Bmat}\uv^\top 
\end{align}
Because the irrelevant variables have been omitted, $\pibar\hat{\Bmat}>0$.
Since $\uv\ge \zerov$ and is not trivially zero, it follows that $\bar{\piv}\hat{\Bmat}\uv^\top>0$. This implies $r(s)<s$. Moreover, if $\epsilon$ is an eigenvalue of $s\Imat+\hat{\Rmat}-\hat{\Dmat}$ then $\epsilon-s$ is an eigenvalue of $\hat{\Rmat}-\hat{\Dmat}$ and has real part
\begin{align}
\real{\epsilon-s} &=\real{\epsilon}-s\nn
&\le \abs{\epsilon}-s\nn
&\le r(s)-s<0.
\end{align}
 Thus the differential equation \eqnref{vvde-matrix-relevant} for the relevant variables is stable and it follows that the differential equation for $\vv(t)$ is also stable. This completes the proof of \Thmref{AOI-SHS}.


\section{Performance Evaluation}\label{sec:perf-eval}
Here we use Theorems~\thmref{ageFCFS} and \thmref{LCFS} to examine achievable AoI regions for two-source FCFS and LCFS systems. In  addition, resource sharing issues for $N$ sources  are explored in Section~\ref{sec:resource-sharing}. 

These issues are pertinent, for example, to embedded systems that are a part of the Internet of Things (IoT)~\cite{munir2015modeling}. As in our models, such systems have multiple sensors (sources), each generating updates independently. These updates are queued for transmission (service) by the system's wireless interface. Their transmission time is modeled by the exponential distribution in this work. Alternatively, the queueing disciplines may be implemented at an access point, which queues updates from a network of a large number of distributed sensors and forwards them to a server for storage and analysis.   


\begin{figure}[t]
\centering
\includegraphics{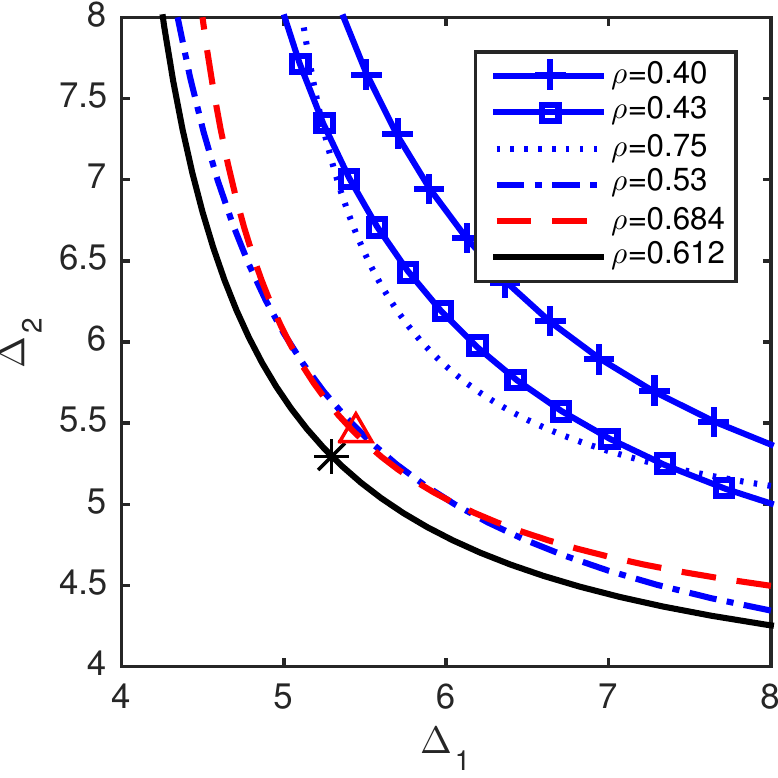}
\caption{Age contours $(\age_1,\age_2)$ for fixed total load $\rho_1+\rho_2=\rho$ for two sources sharing a rate $\mu=1$ FCFS M/M/1 queue. The minimum sum age point marked $*$ in the lower left is achieved by  $\rho_1=\rho_2=0.306$. The point $\triangle$ marks the Nash equilibrium age pair achieved by unilateral optimization.}
\label{fig:FCFSregion} 
\end{figure}

\subsection{M/M/1 FCFS: Two Sources}
For a two-user system with normalized service rate $\mu=1$,  \Thmref{ageFCFS} yields the  contours  of achievable age pairs $(\age_1,\age_2)$ for fixed load $\rho$ that are shown in Figure~\ref{fig:FCFSregion}. The set of feasible age pairs $(\age_1,\age_2)$ is given by the union of all such contours.  
 The lower left  ``corner'' point (marked $*$) where the sum $\age_1+\age_2$ is minimized is obtained at $\rho_1=\rho_2=0.306$, yielding $\age_1=\age_2=5.30$. It then follows from \Thmref{ageFCFS} that if those two sources were to share a rate $\mu=2$ server, then each source would obtain average age  $\age_1=\age_2=2.65$. By comparison, if server resources were partitioned and each source employed an individual rate $\mu=1$ server with optimal load $\rho_1=0.531$, \Eqnref{sysAgeMM1fcfs} will yield  $\age_1=3.48$. Thus we observe a trunking efficiency in having two status-updating sources share a combined service facility.   

\begin{figure}[t]
\centering 
\includegraphics{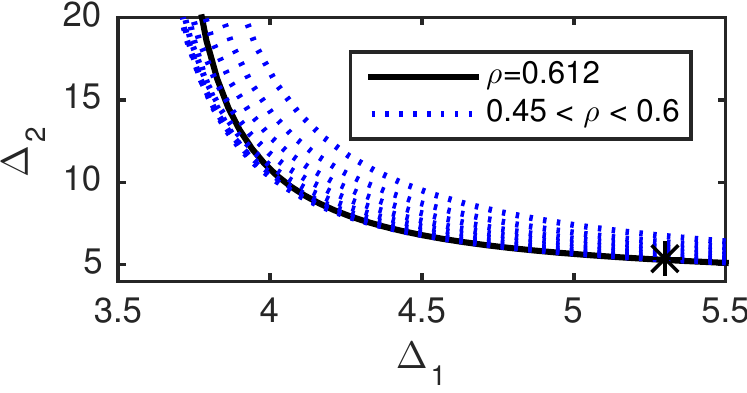}
\caption{Age contours $(\age_1,\age_2)$ for fixed total load $\rho_1+\rho_2=\rho$ for two sources sharing a rate $\mu=1$ FCFS M/M/1 queue. In the lower right corner, the $\rho=0.612$ contour is seen to be Pareto optimal for $\rho_1\approx\rho_2$.  However, in the upper left corner, $\rho\approx 0.53$ can reduce $\age_1$ for constant $\age_2$ as $\rho_2\to0$ and $\rho_1\to\rho.$}
\label{fig:ageplotedgefig} 
\end{figure}

We note that the $\rho=0.612$  age contour appears in Figure~\ref{fig:FCFSregion}  to be superior to all other age contours. In fact, this contour marks the Pareto frontier of achievable ages 
at the operating point $*$, 
corresponding to offered loads $\rho_1=\rho_2$. However, the optimal load $\rho$ will vary along the Pareto frontier. For example, as $\rho_1\to\rho$ and $\rho_2\to0$, the  $\rho=0.53$ (the optimal load for a single source) contour offers reduced $\age_1$ for fixed $\age_2$.
This is shown in Figure~\ref{fig:ageplotedgefig} where the solid line marks the $\rho=0.612$ contour while dotted lines mark contours for $\rho$ in the interval $0.45\le \rho\le 0.6$. The general insight for FCFS systems is that multiuser age optimization depends on both the total load $\rho$ and the allocation of load among individual sources.

\subsection{M/M/1 LCFS: Two Sources}
\begin{figure}[t]
	\centering
	\includegraphics{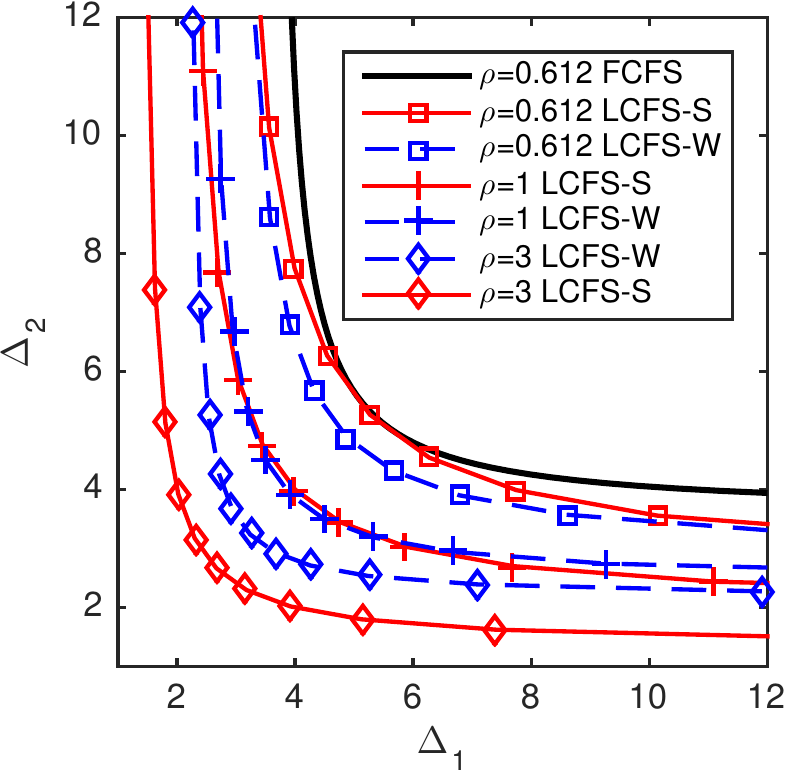}
	\caption{Update ages of two users under the queuing policies of FCFS, LCFS with preemption in service (LCFS-S), and LCFS with preemption only in waiting (LCFS-W). The service rate is $\mu=1$.}
	\label{fig:lcfsfcfsTwoUser}
	\vspace{-2mm}
\end{figure}

With $N=1$ source, some algebra applied to \Thmref{LCFS} will verify that LCFS-S (with preemption in service) yields smaller age $\age_1$ than LCFS-W (with preemption only in waiting). 
For $N=2$ sources, Figure~\ref{fig:lcfsfcfsTwoUser} compares the LCFS policies for different choices of the arrival rate $\rho$. The achievable age pair contour for a given $\rho$ is obtained by varying $\rho_1$ and $\rho_2$ such that $\rho_1 + \rho_2 = \rho$. The service rate is $\mu=1$. We observe that LCFS-W is better than LCFS-S when arrival rates $\rho$ are low but somewhat worse when arrival rates are high.
Because this same behavior does not hold for the single source system, we speculate that LCFS-W benefits at low arrival rates from not preempting an update in service with an update from some other source whose update is not currently in service.

For comparison with FCFS, we also plot the FCFS age contour for total load $\rho=0.612$, which was shown in Figure~\ref{fig:FCFSregion} to be optimal in the neighborhood of $\rho_1\approx\rho_2$ and near-optimal elsewhere. For the same total offered load $\rho=0.612$, Figure~\ref{fig:lcfsfcfsTwoUser} shows that the age contours obtained under both LCFS policies are better than those obtained under FCFS. To summarize Figure~\ref{fig:lcfsfcfsTwoUser}, if a system can choose all of $\rho$, $\rho_1$, and $\rho_2$, LCFS-S is the policy that minimizes sum age.

In general, the choice of policy is not as straightforward. Figure~\ref{fig:lcfsfcfsTwoUserBestPolicy} shows, for $N=2$ sources and $\mu=1$, that the policy that minimizes the sum age over points in the cartesian product of $\{0 < \rho_1/\rho \le 0.5\}$ and $\{0 < \rho < 2\}$. FCFS is the policy of choice for $\rho < 0.4$. It is also the policy of choice for larger $\rho < 1$, however, only when source $1$ updates constitute a small enough fraction of $\rho$. This fraction gets smaller as $\rho \to 1$. Similarly, LCFS-W is a policy of choice over LCFS-S even for large $\rho$ when the load due to source $1$ updates is a small enough fraction of the total load $\rho$.

\subsection{Multiple Source Resource Allocation}\label{sec:resource-sharing}
\begin{figure}[t]
	\centering
	\includegraphics[scale=0.27]{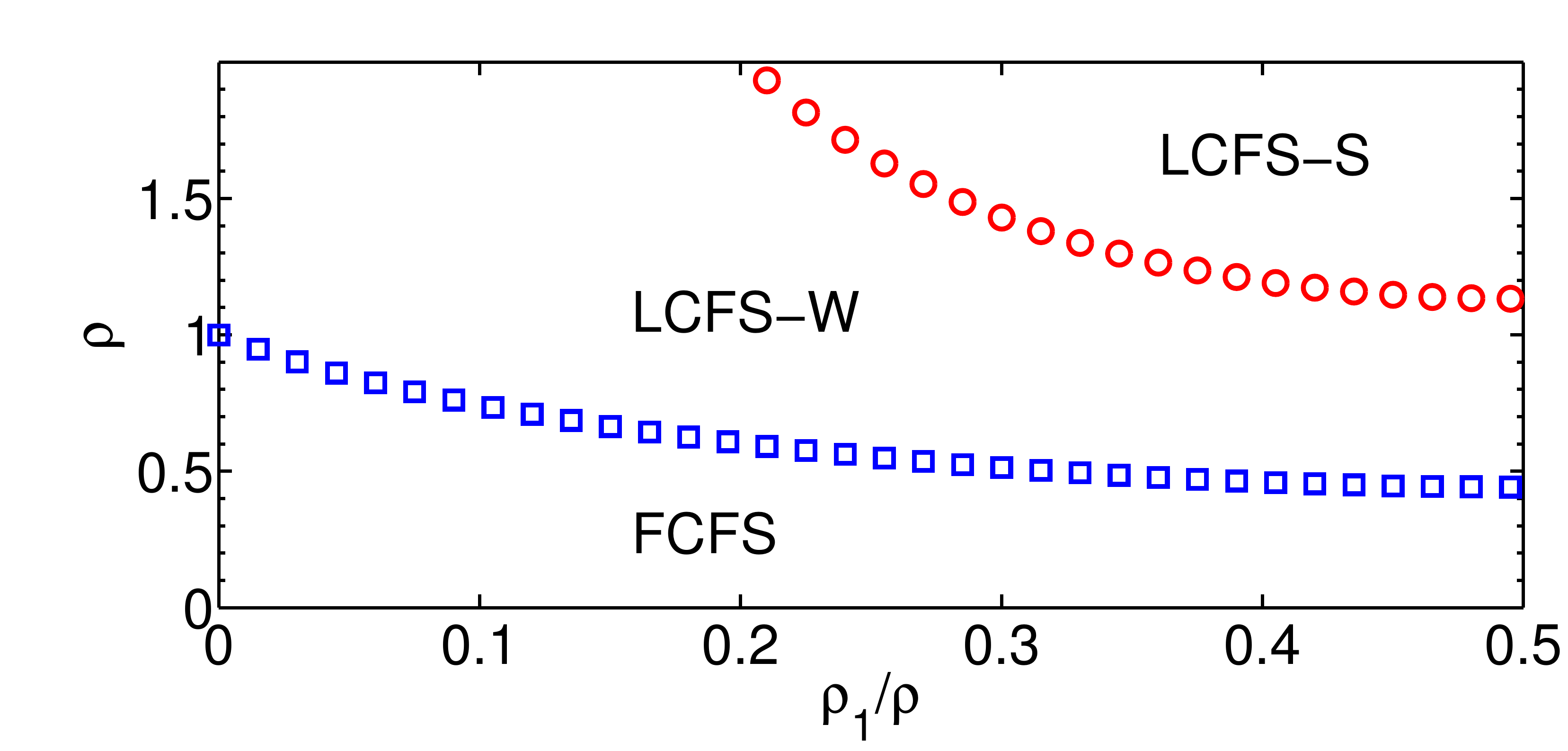}
	\caption{For the policies of FCFS, LCFS-W, and LCFS-S, we show the region where they are sum age minimizing. FCFS minimizes the sum age for all points $(\rho_1/\rho, \rho)$ that lie below the squares. LCFS-W minimizes the sum age for the points above the squares and to the left of the circles. LCFS-S is sum age minimizing for all points above the circles. The service rate is $\mu=1$.}
	\label{fig:lcfsfcfsTwoUserBestPolicy}
\end{figure}

We can compare the FCFS and LCFS systems in terms of the sum of ages $\agesum=\sum_{i=1}^N \age_i$ when users share the system capacity in fixed proportions such that $\rho_i=\alpha_i\rho$ with $\sum_{i=1}^N\alpha_i=1$. We note that \Thmref{LCFS} implies that the following observations hold for both types of LCFS systems:
\begin{itemize}
\item 
Each user $i$ has age $\age_i$ that decreases monotonically with total load $\rho$. 
\item The sum age $\agesum$ is minimized by equal offered loads $\rho_i=\rho/N$.
\end{itemize}
In addition, it follows from \Thmref{LCFS} that
\begin{itemize} 
\item $\age_i$ in a LCFS-W system is strictly less than that under LCFS-S iff
\begin{align}\eqnlabel{Nmax}
\frac{1}{N}\sum_{i=1}^N \alpha_i^{-1} > (1 + \rho) \aw(\rho).
\end{align}
\end{itemize}
The right side of \Eqnref{Nmax} is a nondecreasing  $O(\rho)$ function. Thus given $N$ updaters and the proportions $\alpha_i$ in which they share the load, there is a maximum $\rho$ such that LCFS-W is better than LCFS-S. We had observed this for $N=2$ sources in Figures~\ref{fig:lcfsfcfsTwoUser} and~\ref{fig:lcfsfcfsTwoUserBestPolicy}.

Optimization of the offered load $\rho$ is more complicated in the FCFS system. \Thmref{ageFCFS} implies that for each user $i$, the value of total load $\rho$ that minimizes the average age $\age_i$ depends on the proportional load $\alpha_i$. Load optimization of the FCFS system will depend on a performance metric derived from $\age_1,\ldots,\age_N$. However, it is easy to see that for any given $0 < \rho < 1$, the sum of ages $\agesum$ is minimized when $\alpha_i = 1/N$ for all $i$. This follows from the fact that for a given $\rho$, $\frac{\partial^2 \age_i}{\partial \rho_i^2} > 0$ for $\rho_i \in (0,\rho)$. Also, as a result, the sum of ages is a convex function of $\rho_i$, $i=1,\ldots,N$. Further, all the $\Delta_i$ are the same function over $\rho_i \in (0,\rho)$.

To compare the FCFS, LCFS-S and LCFS-W systems, we focus on symmetric systems with $\rho_i=\rho/N$.  In this case, each user has identical average age $\age_i$ given by \Thmref{ageFCFS} or \Thmref{LCFS} with $\rho_i=\rho/N$ and $\OthersLoad{i}=(N-1)\rho/N$. 
For symmetric LCFS systems, \Eqnref{Nmax} simplifies to
\begin{equation}\eqnlabel{NmaxSym}
N>(1+\rho)\aw(\rho).
\end{equation}
Therefore for fixed $\rho$, LCFS-W outperforms LCFS-S when the number of sources $N$ is large. To further examine this, we now assume a large system with $N\gg1$ sources.  
As $N$ becomes large,  
$\OthersLoad{i}\to\rho$ and it follows from \eqnref{alphaw-limits} that $\aw(\rho)/N\to0$. With these limits, it is straightforward to show that Theorems~\thmref{ageFCFS} and~\thmref{LCFS} imply that $\age_i$ approaches $\Nage_F(\rho)$ (FCFS) or $\Nage_S(\rho)$ (LCFS-S) or $\Nage_W(\rho)$ (LCFS-W) where 
\begin{subequations}\eqnlabel{largeN}
\begin{align}
\NageF(\rho)&=\frac{N}{\mu}\bracket{\frac{(1+\rho)\rho^2}{N^3(1-\rho)^3}
+\frac{1}{N(1-\rho)}+\frac{1}{\rho}}, \eqnlabel{largeN-F}\\
\NageS(\rho)&=\frac{N}{\mu}\bracket{1+\frac{1}{\rho}},\eqnlabel{largeN-S}\\
\NageW(\rho)&=
\frac{N}{\mu}\bracket{1+\frac{1}{\rho(1+\rho)}}.
\eqnlabel{largeN-W}
\end{align}
\end{subequations}
From \eqnref{largeN-S} and \eqnref{largeN-W}, we see that both LCFS systems have AoI that decreases with total load $\rho$. By contrast, the FCFS system  
is subject to the stability constraint $\rho<1$ and it benefits from  matching the load $\rho$ to the number of sources $N$. Let $\rho_N^*$ minimize $\NageF(\rho)$ over $0\le\rho\le1$.  Because $\rho<1$ and $N$ is large, the first term on the right side of \eqnref{largeN-F} becomes negligible for $\rho$ near $\rho_N^*$ as $N$ becomes large. Thus $\rho^*_N$ converges to the minimizer of $[N(1-\rho)]^{-1}+\rho^{-1}$, i.e.,
\begin{equation}
\rho^*_N=\frac{\sqrt{N}}{\sqrt{N}+1}.
\end{equation}
It follows that $\NageF(\rho_N^*)\to N/\mu$ as $N$ becomes large.  Similarly, for the LCFS systems, if $\rho$ grows large as $N$ grows large, then $\NageS(\rho)$ and $\NageW(\rho)$ also approach $N/\mu$.  In this sense, as the number of users becomes large, all three systems become equivalent. On the other hand, in many settings  the offered load $\rho$ may be physically constrained. For example, if each update requires energy for a wireless transmission, then it would be appropriate to compare the FCFS and LCFS systems on an equal power basis. In this spirit, we note that if all three systems operate at offered load $\rho=\rho_N^*$, then $\NageW(\rho^*_N)/\NageF(\rho^*_N) \to 1.5$ and $\NageS(\rho^*_N)/\NageF(\rho^*_N) \to 2$ as $N$ becomes large. Thus,  one can argue that FCFS is more efficient than either LCFS discipline  for large symmetric systems.


\subsection{Non-cooperative Rate Adaptation}
We now examine how sources may individually adapt their updating rates. When source $i$ is a status updater in the presence of ``interfering'' traffic with aggregate load $\OthersLoad{i}$ from other sources, it may be in the interest of source $i$ to unilaterally optimize  its updating load $\rho_i$ in response to the aggregate other load $\OthersLoad{i}$. 

 For the $N$-user FCFS system, it was observed \cite{2012ISIT-YatesKaul} that 
%
the adaptation
\begin{align}\eqnlabel{rho1opt}
\rho^*_i(\OthersLoad{i})=\frac{1-\OthersLoad{i}}{2}+\frac{[1-\OthersLoad{i}]^2}{32}
\end{align}
is essentially the same as the best-response
normalized load that exactly minimizes $\age_i$.
It was also noted that since the minima over $\rho_i$ is broad and nearly flat,  $\rhohat_i=0.5(1-\OthersLoad{i})$,  a rule of thumb that a source should use half the the residual capacity, was a good linear approximation.
If each other source is a status updater that selects  an update load to minimize its respective age given the other sources' update loads, 
 we obtain the synchronous iterative algorithm
 \begin{align}
\rho_i(n+1)&=\frac{1-\OthersLoad{i}(n)}{2}+ \frac{[1-\OthersLoad{i}(n)]^2}{32}. 
\end{align}

This iterative algorithm was shown to work reasonably well for $N=2$ users as it converges to a fixed point with $\rho_1=\rho_2 = 0.342,$ and corresponding ages $\age_1=\age_2=5.4390$ \cite{2012ISIT-YatesKaul}. 
However, it is easily verified that this iteration  is unstable for $N>2$ users. 
Similar distributed algorithms $\rhohat_i(n+1)=\w_N[1-\OthersLoad{i}(n)]$, in which each node uses a fraction $\w_N$ (depending on the number of sources $N$) of the residual capacity can be shown to be stable. However, a network mechanism that sends $\w_N$ to a source $i$ might just as well send the appropriate load $\rho_i$ to that source.  

For both LCFS systems, each source $i$ has incentive to generate updates as fast as possible.
To be precise, when a source $i$ is subject to a maximum load constraint $\rho_i\le \bar{\rho}_i$ and other sources offer combined load $\OthersLoad{i}$, then source $i$ can decrease its average age $\age_i$ by unilaterally increasing $\rho_i$ to $\bar{\rho}_i$. The Nash equilibrium of the system is for each node $i$ to operate at maximum updating load $\bar{\rho}_i$. The average age each node will obtain will depend on $\bar{\rho}_i$ and the  total updating load $\sum_{k=1}^N\bar{\rho}_k$. This may be a desirable operating point for some nodes but not for others. Moreover, if each node bears some cost for its offered load $\rho_i$, this operating point may be undesirable even if the resulting age is small. 

An alternate approach would be for node $i$ to view its age as a function $\age_i(\rho_i,\OthersLoad{i})$ of its offered load and the interfering load, and to adjust $\rho_i$ to meet a target  age constraint $\age_i(\rho_i,\OthersLoad{i})= \delta_i$. Such an approach shares many commonalities with game-theoretic power control in wireless CDMA systems \cite{SaraydarMandayamGoodman2002} although discussion of these issues is beyond the scope of this work.

%

%

\section{Conclusion}
\label{sec:conclusions}
We have looked at the problem of multiple sources generating timely status updates at interested recipients.  We have employed a simple approach in which the communication network is modeled by an M/M/1 queue with either FCFS service or one of two variants of LCFS service.

For these systems, we have derived and characterized the region of feasible status ages. Our results show that there are nontrivial gains in trunking efficiency when $N$ users share the system capacity. However, achieving these gains appears to require coordinated load balancing of the sources. In particular, high offered load at an FCFS system induces high AoI through queueing delays. A lossy LCFS discipline can mitigate this problem but its packet discarding policy may encourage sources to operate at excessively high offered loads.  

The preliminary insights from these simple models lead us to believe that the age of information represents a new and useful metric for the analysis of status updating systems. Moreover, it should be apparent that many challenges remain in modeling, characterizing and optimizing practical status updating systems. 

As a step in this direction, we have introduced stochastic hybrid systems as a new way to evaluate AoI in queues with memoryless service. The SHS method, in the form of \Thmref{AOI-SHS}, provides a way to derive closed-form AoI results for simple queues described by finite-state Markov chains.  In addition,  \Thmref{AOI-SHS} will   permit evaluation of AoI in substantially more complex queueing systems that  capture more realistic service facilities. These include facilities that prioritize service of updates of certain sources over others, heterogeneous service facilities, state-dependent updating policies, and 
facilities in which arrival and service rates change with time.  
More generally, continuous-time Markov chains are a widely used tool for the modeling and performance evaluation of complex  service systems \cite{bolch2006queueing}. When these systems are delivering status updates, \Thmref{AOI-SHS} enables straightforward numerical evaluation of AoI. 

\Thmref{AOI-SHS} is based on non-negative linear transition reset maps. We note that this choice is not unique and that there are other ways to embed AoI tracking in the continuous state of an SHS with linear reset maps. However, stability results for the corresponding differential equations will need to be derived.  We note that there is substantial literature on SHS stability and ergodicity \cite{teel2014stability}, but this is not yet fully understood in the specific context of age systems. We further observe that a general SHS with time-varying and state-dependent transition rates also may prove to be useful in further characterizing age processes.

\appendices
\section{Proof of Lemma~\ref{lem:EWX}}
\label{sec:EWXproof}
The proof of Lemma~\ref{lem:EWX} relies on the following basic properties of Poisson processes and exponential random variables.
\begin{lemma}\label{lem:expos} Let $X_1$ and $X_2$ be independent exponential random variables with $\E{X_i}=1/\alpha_i$. Let $V=X_2-X_1$.  
\begin{letterate}
\item $\prob{X_1<X_2}=\alpha_1/(\alpha_1+\alpha_2)$.
\item Given $X_1<X_2$, $X_1$ and $V$ are conditionally independent and have conditional exponential 
probability density functions (PDFs)
\begin{align*}
\pdf{X_1|X_1<X_2}{x} &=
(\alpha_1+\alpha_2)e^{-(\alpha_1+\alpha_2)x},& x\ge 0,\\
\pdf{V|X_1<X_2}{v}&=\alpha_2e^{-\alpha_2v},& v\ge 0.
\end{align*} 
\end{letterate}
\end{lemma}

\begin{lemma}\label{lem:geom}
Given a rate $\lambda$ Poisson process $N(t)$ and an independent exponential $(\alpha)$  random variable $X$, the number of arrivals $N(X)$ in the interval $[0,X]$ has the geometric PMF
$$\pmf{N(X)}{n}=
(1-\gamma)\gamma^n,\qquad  n\ge 0,\\
$$
with $\gamma=\lambda/(\alpha+\lambda)$.
 \end{lemma}

To prove Lemma~\ref{lem:EWX}, let $Y_j$, $W_j$ and $T_j$ to refer to the interarrival  time, waiting time and system time of the $j$th packet of source $i$. We now examine 
$W_j$ via the partition 
\begin{align}
\jbrief&=\set{Y_j<T_{j-1}},\quad 
\jlong =\set{T_{j-1}< Y_{j}}.
\end{align}
That is, $\jbrief$ denotes the event that the $j$th interarrival time for source $i$ is brief, specifically, less than the system time of the preceding packet from source $i$, 
and $\jlong$ is the complementary event  that $Y_{j}$ is long. 
With the partition $\set{\jbrief,\jlong}$, we write
\begin{align}
\E{Y_{j}W_{j}}&=
\E{Y_{j}W_{j}|\jlong}\prob{\jlong}
+\E{Y_{j}W_{j}|\jbrief}\prob{\jbrief}.
\eqnlabel{EWX-2cases}
\end{align}
Since source $i$ packets and  packets from other sources have identical exponential $(\mu)$ service times, the combined queue is just M/M/1 with offered load $\rho=\rho_i+\OthersLoad{i}$. In steady state, the system time
$T_{j-1}$ has the exponential $(\mu-\lambda)$ PDF 
\begin{align}
\ipdf{T} &= (\mu-\lambda) e^{-(\mu-\lambda)t},\qquad  t\ge 0.
\eqnlabel{pdfTmm1}
\end{align}
Furthermore, $T_{j-1}$, which depends on packets (and their service times) that arrived prior to packet $j-1$, 
is independent of $Y_{j}$. 
Given $\jbrief$,  packet $j-1$ is still in the system when packet $j$ is generated. The waiting time $W_{j}$ depends on both the residual system time $T_{j-1}-Y_{j}$ and also on the workloads of packets from other sources that arrive during the source $i$ interarrival period of length $Y_{j}$.  Specifically, let $M=N_{-i}(Y_{j})$ denote the number of other-source (i.e. not source $i$) packets that arrive during the source $i$ interarrival period and let $S_{1},S_{2},\ldots S_{M}$ denote their service requirements. As these  packets are all queued between packets $j-1$ and $j$, 
\begin{align}
W_{j} = (T_{j-1}-Y_{j})+ \sum_{k=1}^{M}S_{k}.
\end{align}
This implies 
$\E{Y_{j}W_{j}|\jbrief} 
=E_1+E_2$
where
\begin{subequations}
\eqnlabel{E12defn}
\begin{align}
E_1&=\E{Y_{j}(T_{j-1}-Y_{j})|\jbrief},\\
E_2&=\E[\bigg]{Y_{j}\sum_{k=1}^{M}S_{k}|\jbrief}. 
\end{align}
\end{subequations}
By Lemma~\ref{lem:expos}(b), 
\begin{align}
E_1 
 &=\E{(T_{j-1}-Y_{j})|\jbrief} \E{Y_{j}|\jbrief}\nn
&=\frac{1}{\mu-\lambda}\parfrac{1}{\lambda_i+ (\mu-\lambda)}
=\frac{1}{\mu^2(1-\rho)(1-\OthersLoad{i})}.\eqnlabel{E1final}
 \end{align}
For the second term, iterated expectation
yields
\begin{align}
E_2 
&=\int_0^\infty \E[\bigg]{Y_{j}\sum_{k=1}^{M}S_{k}|\jbrief, Y_{j}=y}\pdf{Y_{j}|\jbrief}{y}\,dy\nn
&=\int_0^\infty \E[\bigg]{y\sum_{k=1}^{M}S_{k}|Y_{j}=y}\pdf{Y_{j}|\jbrief}{y}\,dy.
\end{align}
Given that $Y_{j}=y$, $M=N_{-i}(Y_{j})=N_{-i}(y)$ is the number of other-source arrivals in a period of length $y$ and is Poisson with conditional expectation $\E{M|Y_{j}=y}=\OthersRate{i}y$. In addition, each $S_{k}$ is an exponential $(\mu)$ random variable, independent of $Y_{j}$, implying $\E{S_{k}|Y_{j}=y}=1/\mu$. 
This implies
\begin{align}
\E[\bigg]{y\sum_{k=1}^{M}S_{k}|Y_{j}=y}
&= y \E{M|Y_{j}=y} \E{S_{k}|Y_{j}=y}
=
y(\OthersRate{i} y)(1/\mu)= \OthersLoad{i} y^2.
\end{align}
By Lemma~\ref{lem:expos}, $Y_{j}$ given $\jbrief$ is an exponential $(\alpha)$ random variable with 
$\alpha= \lambda_i+(\mu-\lambda_i-\OthersRate{i})=\mu-\OthersRate{i}.$
This implies
\begin{align}
E_2 &=\OthersLoad{i}\int_0^\infty y^2 \alpha e^{-\alpha y}\,dy
=\frac{2\OthersLoad{i}}{\alpha^2}=\frac{2\OthersLoad{i}}{\mu^2(1-\OthersLoad{i})^2}.\eqnlabel{E2final}
\end{align}
It follows from 
\eqnref{E1final} and  \eqnref{E2final} that
\begin{align}\eqnlabel{EWX2-XltT}
\E{{Y_{j}W_{j}}|\jbrief}&=\frac{1}{\mu^2}
\bracket{\frac{2\OthersLoad{i}}{(1-\OthersLoad{i})^2}+\frac{1}{(1-\OthersLoad{i})(1-\rho)}}.
\end{align}

Given event $\jlong$,  packet $j-1$ has departed the system prior to the arrival of packet $j$. In this case, the waiting time for packet $j$ depends on the number of other-source  packets in the system when packet $j$ arrives. To characterize this, we now let $M$ denote the number of other-source packets in the system  at the departure instant of packet $j-1$. Since the queue is FCFS, $M$ is the number of other-source packets that arrived and were queued during the system time $T_{j-1}$ of packet $j-1$. Given $T_{j-1}$ is exponential and independent of $Y_{j}$, Lemma~\ref{lem:expos}(b) tells us that $T_{j-1}$ given $\jlong$ is conditionally an exponential $(\alpha)$ random variable with   $\alpha= (\mu-\lambda)+\lambda_i=\mu-\OthersRate{i}$. As $T_{j-1}$ is independent of the subsequent Poisson  arrivals of the other sources, Lemma~\ref{lem:geom} implies that $M$ has the geometric PMF
\begin{align}\eqnlabel{M2dist}
\pmf{M}{m}=(1-\gamma)\gamma^m,\qquad m\ge0,
\end{align}
where  $\gamma ={\OthersRate{i}}/({\alpha+\OthersRate{i}})
=\OthersLoad{i}$.

From \eqnref{M2dist}, we see  at the departure instant of packet $j-1$  
that $M$ is described by the stationary distribution of an M/M/1 queue serving only other-source packets at rate $\OthersRate{i}$. Going forward from this instant, we wait an additional time $Y_{j}-T_{j-1}$ for packet $j$ from source $i$. In this time period, there may be either arrivals or departures of other-source packets. Nevertheless, as the queue holds zero source $i$ packets, the operation of the queue is identical to an M/M/1 queue for just other-source packets. At all times up to the arrival of packet $j$, the number of other-source packets in the queue remains stochastically identical to $M$. If the $k$th queued other-source packet has service requirement $S_{k}$, then 
$W_{j}=\sum_{k=1}^{M}S_{k}$ and $\E{W_{j}|L_j}=\E{M}/\mu$.
It follows that when packet $j$ does arrive, the number of queued packets $M$ and the service times $S_k$ are  independent of both the additional delay $Y_{j}-T_{j-1}$ until the arrival of packet $j$ and the  prior system time $T_{j-1}$. This implies
\begin{align}
\E{Y_{j}W_{j}|\jlong}
&=\E{(T_{j-1}+(Y_{j}-T_{j-1}))W_{j}|\jlong}\nn
&=\E{T_{j-1}+(Y_{j}-T_{j-1})|\jlong}\E{W_{j}|\jlong}\nn
&=\paren{\frac{1}{\mu-\OthersRate{i}}+\frac{1}{\lambda_i}}
\parfrac{\OthersLoad{i}}{\mu(1-\OthersLoad{i})}.\eqnlabel{EWX2-TltX}
\end{align}
Next we recall from Lemma~\ref{lem:expos} that independence of $T_{j-1}$ and $Y_{j}$ implies $\prob{\jbrief}={\rho_i}/({1-\OthersLoad{i}})$.
Combining this fact with \eqnref{EWX-2cases}, \eqnref{EWX2-XltT}, and  \eqnref{EWX2-TltX},
some algebra yields the claim.
\section{Theorem~\ref{thm:LCFS}(a): Proof Completion}
\label{sec:Proof:LCFS-S}
Note that $\Spre=\Spre_j$ is the time that packet $j$ from source $i$ spends in service. This packet completes service (is not preempted) and hence no new arrivals occur during the interval
$S_L$. 
The distribution of $\Spre$ is that of the time to service completion, say $X_\mu$, after packet $j$ arrives, conditioned on $X_\mu$ being smaller than the time to the next packet arrival, say $X_\lambda$, from any source. Thus $P[\Spre > z] = P[X_\mu > z|X_\mu < X_\lambda]$. By Lemma~\ref{lem:expos}(b), $\Spre_j$ is exponential $(\lambda+\mu)$ and 
$\E{\Spre} = {1}/({\lambda + \mu})$.
%

Now we calculate the moments of $\interDep=\interDep_j$. From~(\ref{eqn:2userB}) we know that $\interDep$ is a random sum of random variables $\interDblock_k$, $1\le k\le L$. Also, $\interDep$ ends with the departure of a source $i$ update packet. Since the arrival rate $\lambda$ is the sum of rates of independent Poisson sources, the probability that any block $\interDblock_k$ ends in the departure of the update packet of source $i$ is $\lambda_i/\lambda$. Note that $\interDep$ consists of $L=l$ blocks if $l-1$ consecutive blocks end in departures of other-source packets, followed by block $l$  ending in a source $i$ departure. Thus,
\begin{align}
\prob{L=l} = (1-q)^{l-1}q, \qquad l\ge1,
\label{eqn:2userPL}
\end{align}
where $q = \lambda_i/\lambda$.
Note that $\interDblock_k = X'_k + S_k$, where $X'_k$ is an idle period and $S_k$ is the server busy period. During the busy period a random number of packets in service 
may be preempted, but the service rate (i.e., the instantaneous departure rate)  is $\mu$ throughout the busy period.
Thus the busy period $S_k$ is memoryless 
and is independent of the number of arrivals during it that get preempted and the source whose packet departs at its end. From these observations and given Poisson arrivals of rate $\lambda$, we can write
\begin{align}
\E{X'_k} = \frac{1}{\lambda},\ \E{S_k} = \frac{1}{\mu},\text{ and } \E{\interDblock_k} = \frac{1}{\lambda} + \frac{1}{\mu}.
\label{eqn:2userECi}
\end{align}
The memoryless nature of the arrival and service processes also implies that each $\interDblock_k$ is independent of $L$. Using this fact and equations~(\ref{eqn:2userB}),~(\ref{eqn:2userPL}) and~(\ref{eqn:2userECi}), we can write
\begin{align}
\E{\interDep} = \E{L}\E{\interDblock_k} = \frac{\mu + \lambda}{\lambda_i\mu}.
\label{eqn:2userEY-a}
\end{align}
To calculate $\E{\interDep^2}$,  let the random variable $\interDblock$ be stochastically identical to block lengths $\interDblock_k$, $k=1,\ldots,L$. Using arguments we used to calculate $\E{\interDep}$, and noting that $\interDblock_k$ and $\interDblock_{k'}$ are independent for $k\ne k'$, we can write
\begin{align}
\E{\interDep^2} = \E{L}\E{\interDblock^2} + \E{L(L-1)}(\E{\interDblock})^2.
\label{eqn:2userEB2(a)}
\end{align}
Also note that the idle period $X'_k$ and busy period $S_k$ that constitute $\interDblock_k$ are mutually independent. This implies
\begin{align}
\E{\interDblock^2} = \frac{2}{\lambda^2} + \frac{2}{\mu^2} + \frac{2}{\lambda\mu}.
\label{eqn:2UserEC2}
\end{align}
Using equations~(\ref{eqn:2userPL}),~(\ref{eqn:2userECi}) and~(\ref{eqn:2UserEC2}), we can write~(\ref{eqn:2userEB2(a)}) as
\begin{align}
\E{\interDep^2} = 2\frac{\lambda}{\lambda_i}\paren{\frac{\lambda}{\lambda_i}\left[\frac{1}{\lambda}+\frac{1}{\mu}\right]^2-\frac{1}{\lambda\mu}}.
\label{eqn:2userEB2}
\end{align}
Applying  the moments $\E{\Spre}$, $\E{\interDep}$ and $\E{\interDep^2}$
to equation~(\ref{eqn:age_lcfs_with_p}) yields \Thmref{LCFS}(a).
\section{Proof of Lemma~\ref{lem:pi-vv-derivs}}
\label{sec:lem:pi-vv-derivs}
From \eqnref{Lpsi-defn},  we calculate $\Lpsim[]{\qbar}$ and $\Lpsim{\qbar}$ for each $j$ and $\qbar\in\Qcal$:
\begin{subequations}\eqnlabel{Lpsim-S-defn}
\begin{align}
\Lpsim[]{\qbar}(q,\xv)&=\Lamm[]{\qbar}(q,\xv),\eqnlabel{Lpsim-S-defn1}\\
\Lpsim{\qbar}(q,\xv)&= \bracket{\bv_q}_j\dq{\qbar}+ \Lamm{\qbar}(q,\xv),\ j\in\range{n},\eqnlabel{Lpsim-S-defn2}
\end{align}
\end{subequations}
where 
\begin{subequations}\eqnlabel{Lammiqbar}
\begin{align}
\Lamm[]{\qbar}(q,\xv) &= \sum_{l\in\Lcal}
\bracket{\psim[]{\qbar}(\phi_l(q,\xv))
-\psim[]{\qbar}(q,\xv)}\laml(q),\\
\Lamm{\qbar}(q,\xv)&=\sum_{l\in\Lcal}
\bracket{\psim{\qbar}(\phi_l(q,\xv))
-\psim{\qbar}(q,\xv)}\laml(q).
\end{align}
\end{subequations}

When $\phi_l(q,\xv)=(\ql',\xv\Amat_l)$,
\begin{subequations}\eqnlabel{psim1}
\begin{align}
\psim[]{\qbar}(\phi_l(q,\xv))&=\psim[]{\qbar}(\ql',\xv\Amat_l) = \dq[\ql']{\qbar},\\
\psim{\qbar}(\phi_l(q,\xv))&=\psim{\qbar}(\ql',\xv\Amat_l)
=\cvec{\xv\Amat_l}_j\dq[\ql']{\qbar},\ j\in\range{n}.
\end{align}
\end{subequations}

%
Since $\laml(q)=\laml\dq{\ql}$, it follows from \eqnref{Lammiqbar} and \eqnref{psim1} that
\begin{subequations}\eqnlabel{Lamm-S-defn}
\begin{align}
\Lamm[]{\qbar}(q,\xv)&=\sum_{l\in\Lcal}\laml\bracket{\dq[\ql']{\qbar}-\dq{\qbar}}\dq{\ql},\\
\Lamm{\qbar}(q,\xv)&=\sum_{l\in\Lcal}\laml\bracket{[\xv\Amat_l]_j\dq[\ql']{\qbar}-x_j\dq{\qbar}}\dq{\ql},\ j\in\range{n}.
\end{align}
\end{subequations}


We observe that
 \begin{subequations}\eqnlabel{kronecker-sift}
\begin{align}
\dq[\ql']{\qbar}\dq{\ql}&=\begin{cases}
\dq{\ql} & l \in \Lcal'_{\qbar},\\
0 & \ow,
\end{cases}\\
\dq{\qbar}\dq{\ql} &=\begin{cases}
\dq{\qbar} & l\in\Lcal_{\qbar},\\
0 & \ow.
\end{cases}
\end{align}
\end{subequations}
It follows from \eqnref{Lcalqbar}, \eqnref{Lamm-S-defn} and \eqnref{kronecker-sift} that
\begin{subequations}\eqnlabel{Lammlast}
\begin{align}
\Lamm[]{\qbar}(q,\xv)&=\sum_{l\in\Lcal'_{\qbar}}\laml\dq{\ql}-\dq{\qbar}\sum_{l\in\Lcal_{\qbar}}\laml,\eqnlabel{Lammlast-pi}\\
\Lamm{\qbar}(q,\xv)&=\sum_{l\in\Lcal'_{\qbar}}\laml {[\xv\Amat_l]_j}
\dq{\ql}-x_j\dq{\qbar}\sum_{l\in\Lcal_{\qbar}}\laml,\ j\in\range{n}.
\eqnlabel{Lammlast-j}
\end{align}
\end{subequations}

With $\psi(q,\xv)=\psim[]{\qbar}(q,\xv)$,  \eqnref{piqhat-defn}, \eqnref{dynkins} and \eqnref{Lammlast-pi} imply
\begin{align}
\dot{\pi}_{\qbar}(t)&=\E{\Lamm[]{\qbar}(q(t),\xv(t))}\nn
&=\E{\sum_{l\in\Lcal'_{\qbar}}\laml\dqt{\ql}-\dqt{\qbar}\sum_{l\in\Lcal_{\qbar}}\laml}\nn
&=
\sum_{l\in\Lcal'_{\qbar}}\laml
\pi_{\ql}(t)- \pi_{\qbar}(t)\sum_{l\in\Lcal_{\qbar}}\laml,\qquad \qbar\in\Qcal.\eqnlabel{statprobs}
\end{align}

For $j\in \range{n}$ with $\psi(q,\xv)=\psim{\qbar}(q,\xv)$,  \eqnref{vqi-defn}, \eqnref{dynkins} and \eqnref{Lammlast-j} imply
\begin{align}
\dot{v}_{\qbar j}(t)&=\E{\Lpsim{\qbar}(q(t),\xv(t))}\nn
&=\E{\bracket{\bv_{q(t)}}_j\dqt{\qbar}}
+\E{\Lamm{\qbar}(q(t),\xv(t))}.\eqnlabel{viqde1}
\end{align}
From \eqnref{Lammlast-j}, we observe that
\begin{align}
\E{\Lamm{\qbar}(q(t),\xv(t))}
&=\sum_{l\in\Lcal'_{\qbar}}\laml 
\bracket{\E{\xv(t)\dqt{\ql}}\Amat_l}_j-\E{x_j(t)\dqt{\qbar}}\sum_{l\in\Lcal_{\qbar}}\laml\nn
&=\sum_{l\in\Lcal'_{\qbar}}\laml\bracket{\vv_{\ql}(t)\Amat_l}_j
-v_{\qbar j}(t)\sum_{l\in\Lcal_{\qbar}}\laml.\eqnlabel{Lammlast2}
\end{align}
It then follows from \eqnref{viqde1} and \eqnref{Lammlast2} that for $j>0$,
\begin{align}
\dot{v}_{\qbar j}(t)
&=
\bracket{\bv_{\qbar}}_j\pi_{\qbar}(t)+\sum_{l\in\Lcal'_{\qbar}}\laml \bracket{\vv_{\ql}(t)\Amat_l}_j
-v_{\qbar j}(t)\sum_{l\in\Lcal_{\qbar}}\laml.\eqnlabel{viqde2}
\end{align}
 The claim for $\dot{\vv}_{\qbar}(t)$ follows by
gathering the equations \eqnref{viqde2} for $j\in\range{n}$ and rewriting as row vectors.


%
\IEEEpeerreviewmaketitle

\begin{spacing}{1.0}
\bibliographystyle{IEEEtran}
\bibliography{paper,ry-it}
\end{spacing}

\end{document}